\documentclass[12pt]{amsart}
\allowdisplaybreaks

\usepackage[T1]{fontenc} 

\usepackage{amssymb,amsfonts} 
\usepackage[mathscr]{eucal}
\usepackage{multicol}
\usepackage{hyperref}

\usepackage{graphicx}

\newcommand{\captionfont}[1]{\textit{\textbf{#1}}} 
 
\newcommand{\opaV}{\mathscr{V}}
\newcommand{\opaK}{\mathscr{K}} 
\newcommand{\voaK}{\mathfrak{K}}
\newcommand{\voaV}{\mathfrak{V}} 
\newcommand{\opaW}{\mathscr{W}}
\newcommand{\modW}{\mathfrak{W}}  
\newcommand{\modL}{\mathscr{L}}   
\newcommand{\modM}{\mathscr{M}}   
\newcommand{\modX}{\mathscr{X}}   

\newcommand{\Hom}{\mathop{\mathrm{Hom}}}

\newcommand{\Oneshow}{\boldsymbol{1}}

\newcommand{\Qshow}{\boldsymbol{Q}}
\newcommand{\Hshow}{\boldsymbol{H}} 

\def\aff           {[[z,z^{-1}]]}
\def\alg           {algebra}
\def\alph          {Q} 
\def\AMS           {{American Mathematical Society}}
\def\cA            {\mathcal{A}}
\def\cC            {\mathcal{C}}

\def\cft           {conformal field theory}
\def\cfts          {conformal field theories}

\def\cO            {\mathcal{O}}
\def\con           {conformal }
\def\Con           {Conformal }

\def\cS            {\mathscr{S}}

\def\dd            {\partial}

\def\End           {\mathop{\mathrm{End}}}
\def\findim        {fi\-ni\-te-di\-men\-si\-o\-nal}

\def\half          {\frac12}

\newcommand\hSL[1] {\widehat{\sL}(#1)}
 
\def\id            {\mathop{\mathrm{id}}\nolimits}
\newcommand\ket[1] {|#1\rangle}
\newcommand\Ket[1] {#1}
\def\ketinfty      {\mho}
\def\lcft          {logarithmic \cft}

\newcommand\N[1]   {N\,{=}\,#1}
\newcommand{\oC}   {\mathbb{C}}
\newcommand{\oQ}   {\mathbb{Q}}
\def\oN            {\mathbb{N}}
\def\oP            {\mathbb{P}}
\def\one           {\id_K}
\def\One           {\mathbf{1}}
\def\ope           {operator product expansion}
\def\opepole[#1][#2,#3]{\left[{#2},{#3}\right]^{\phantom{y}}_{#1}}
\def\logopepole[#1][#2,#3]{\left[{#2},{#3}\right]^{*}_{#1}}
\def\oZ            {\mathbb{Z}} 
\def\p             {\partial}
\def\rep           {representation}
\def\rme           {\mathrm{e}}

\def\sL            {{\mathfrak{sl}}}
\newcommand\SL[1]  {\sL(#1)}

\def\talgO         {\widetilde{\opaV}}
\def\tensor        {\mathbin{\otimes}}
\def\Tensor        {\,{\otimes}\,}
\renewcommand\tfrac[2]{\mbox{\large$\frac{#1}{#2}$}}
\def\thalf         {\tfrac12}
\def\tilde         {\widetilde}
\def\Vac           {\Omega}
\def\vac           {\omega}
\def\vfi           {\varphi} 
\def\voa           {vertex operator algebra}

\newtheorem{Thm}{Theorem}[section]
\newtheorem{Lemma}[Thm]{Lemma}

\theoremstyle{definition}

\newtheorem{Rem}[Thm]{Remark}

\catcode`\@=11

\def\citen#1{\if@filesw \immediate\write \@auxout {\string\citation{#1}}\fi%
\@tempcntb\m@ne \let\@h@ld\relax \def\@citea{}%
\@for \@citeb:=#1\do {\@ifundefined {b@\@citeb}%
    {\@h@ld\@citea\@tempcntb\m@ne{\bf ?}%
    \@warning {Citation `\@citeb ' on page \thepage \space undefined}}%
    {\@tempcnta\@tempcntb \advance\@tempcnta\@ne
    \setbox\z@\hbox\bgroup\ifcat0\csname b@\@citeb \endcsname \relax
    \egroup \@tempcntb\number\csname b@\@citeb \endcsname \relax
    \else \egroup \@tempcntb\m@ne \fi \ifnum\@tempcnta=\@tempcntb
    \ifx\@h@ld\relax \edef \@h@ld{\@citea\csname b@\@citeb\endcsname}%
    \else \edef\@h@ld{\hbox{--}\penalty\@highpenalty
    \csname b@\@citeb\endcsname}\fi
    \else \@h@ld\@citea\csname b@\@citeb \endcsname \let\@h@ld\relax \fi}%
\def\@citea{,\penalty\@highpenalty\hskip.13em plus.13em minus.13em}}\@h@ld}
\def\@citex[#1]#2{\@cite{\citen{#2}}{#1}}%
\def\@cite#1#2{\leavevmode\unskip\ifnum\lastpenalty=\z@\penalty\@highpenalty\fi%
  \ [{\multiply\@highpenalty 3 #1%
  \if@tempswa,\penalty\@highpenalty\ #2\fi}]}   %

\@addtoreset{equation}{section}
\def\@secnumfont{\bfseries}
\def\subsubsection{\@startsection{subsubsection}{3}%
  \z@{.5\linespacing\@plus.7\linespacing}{-.5em}%
  {\normalfont\bfseries}}
\def\paragraph{\@startsection{paragraph}{4}%
  \z@\z@{-\fontdimen2\font}%
  \normalfont\bfseries}
\def\subparagraph{\@startsection{subparagraph}{5}%
  \z@\z@{-\fontdimen2\font}%
  \normalfont\bfseries}

\catcode`\@=12

\newcommand\wb{\,\linebreak[0]} \def\wB{$\,$\wb}
\def\Bi              {\bibitem}
\newcommand\J[5]     {{\sl #5}, {#1} {#2} ({#3}) {#4}}
\newcommand\Dipl[2]  {{\sl #2}, Diploma thesis (#1)}
\newcommand\PhD[2]   {{\sl #2}, Ph.D.\ thesis (#1)}
\newcommand\Prep[2]  {{\sl #2}, preprint {#1}}
\newcommand\BOOK[4]  {{\em #1\/} ({#2}, {#3} {#4})}

\def\comp  {Com\-mun.\wb Math.\wb Phys.}
\def\fuaa  {Funct.\wb Anal.\wb Appl.}
\def\jhep  {J.\wb High\wB Energy\wB Phys.}
\def\mpla  {Mod.\wb Phys.\wb Lett.\ A}
\def\nupb  {Nucl.\wb Phys.\ B}
\def\ijmp  {Int.\wb J.\wb Mod.\wb Phys.\ A}
\def\mams  {Memoirs\wB Amer.\wb Math.\wb Soc.}

\def\phlb  {Phys.\wb Lett.\ B} 

\begin{document}


\title[Logarithmic Deformations]{\mbox{}\hfill
  \texttt{\lowercase{hep-th/0201091}}\\[1.5\baselineskip]
  Logarithmic Conformal Field Theories\\[2mm]
  via Logarithmic Deformations}

\author[Fjelstad]{J.~Fjelstad}

\author[Fuchs]{J.~Fuchs} 

\author[Hwang]{S.~Hwang} 
\address{JFj, JF, SH: Karlstad University, Karlstad, Sweden}

\author[Semikhatov]{A.M.~Semikhatov} 
\address{AMS, IYT: Lebedev Institute, Moscow, Russia}

\author[Tipunin]{I.Yu.~Tipunin}


\begin{abstract}
  We construct logarithmic conformal field theories starting from an
  ordinary conformal field theory\,---\,with a chiral algebra $\cC$
  and the corresponding space of states $V$\,---\,via a two-step
  construction: i)~deforming the chiral algebra representation on
  $V\Tensor\End K\aff$, where $K$ is an auxiliary finite-dimensional
  vector space, and \ ii)~extending $\cC$ by operators corresponding
  to the endomorphisms $\End K$.  \ For $K\,{=}\,\oC^2$, with $\End K$
  being the two-dimensional Clifford algebra, our construction results
  in extending $\cC$ by an operator that can be thought of as
  $\dd^{-1}\!E$, where $\oint\!  E$ is a fermionic screening.  This
  covers the $(2,p)$ Virasoro minimal models as well as the $\hSL2$
  WZW theory.
\end{abstract}

\maketitle

\thispagestyle{empty}

\flushcolumns
\setcounter{tocdepth}{2}

\vspace*{-40pt}

\begin{center}
  \emergencystretch12pt
 \parbox{.95\textwidth}{
   \begin{multicols}{2}
     {\footnotesize
       \tableofcontents}
   \end{multicols}
   }
\end{center}


\section{Introduction}
\label{sec:introduction}
It is somewhat miraculous that logarithmic conformal field
theories~\cite{gura,gaKa,floh5,raak,kole,gaKa3,morS,floh11}, although
violating some of the fundamental principles usually assumed in CFT,
nevertheless emerge in many situations where conformal symmetry plays
a role.  Applications that have recently been attracting attention
reach from gravitational dressing \cite{biko}, extended symmetries in
disordered systems, polymers and
percolation~\cite{sale9,cakt,gura2,card16,gulu} and WZW models on
supergroups~\cite{rosa,bhas,ludw4,koNI} to the D-brane
dynamics in string theory~\cite{komw,masz2} and the AdS/CFT
correspondence~\cite{ghkA,myle,koga2,sanj,giri,koPo,morS3}.  On the
representation-theoretic side, logarithmic conformal field theories
are of interest because studying nondiagonalizable Virasoro
representations~\cite{rohs,rohs2} may require an extension (and
certainly, new applications) of the ideas and methods that have been
successfully used in constructing mathematical foundations of the
``standard'' conformal field theory.

Although there is no precise definition of the class of logarithmic
conformal field theories -- we comment on this issue at the end of
Sec.~\ref{sec:conclusions} -- the two characteristic properties are
the nondiagonalizable Virasoro action and the appearance of logarithms
in correlation functions (these issues are closely related; see,
e.g.,~\cite{mila3} on how the maximum Jordan cell size restricts the
maximal power of logarithms).  A trivial example of logarithms
occurring in correlation functions is provided by the conformal field
theory of a free boson.  However, the logarithm in this case merely
corresponds to including the position as an operator (conjugate to
momentum) on the state space, and this inclusion does not lead to any
interesting new structure for the \rep s of the relevant chiral
algebra (the $\mathfrak u(1)$ current algebra).  On the other hand, in
``genuinely'' logarithmic models, the logarithms are closely tied to
reducible, but indecomposable modules over the
Virasoro~\cite{rohs,rohs2} or extended chiral algebras.  In such
modules, the action of $L_0$ is not diagonalizable, but rather
connects vectors in a Jordan cell of a certain size.

In a number of examples studied in the literature, Jordan cells occur
at the top level of the module, which in the case of the vacuum module
implies the existence of a ``logarithmic partner to the identity
operator.'' However, this behavior is not generic.  Moreover,
logarithmic partners of the energy-momentum tensor need not
necessarily appear (although their existence seems to have sometimes
been taken nearly as the definition of a logarithmic conformal field
theory).  It was indeed already noticed in~\cite{rohs2} that Jordan
cells can first occur at lower levels, giving rise to the so-called
staggered modules, in contrast to Jordan highest-weight
modules~\cite{rohs2}, where Jordan cells are ``inherited'' from the
top level of the module.\footnote{~The case studied in~\cite{rohs2} is
  that of Jordan \textit{Verma} modules\,---\,the ``logarithmic
  version'' of Verma modules, while what occurs in many specific
  models are not Verma, but Fock (or rather, Feigin--Fuks
  (FF)~\cite{fefu}) modules.  Accordingly, the notions introduced
  in~\cite{rohs2} should be applied modulo the corresponding
  ``correction.''  Compared to Verma modules, the FF modules are
  distorted by inverting ``one half'' of the embeddings.  This
  relationship carries over to their logarithmic versions.}  The
``staggered'' case is in fact most frequent in applications, in
particular in logarithmically extended minimal models.

In this paper, we present a construction of logarithmic conformal
field theories via a certain deformation of conventional conformal
field theories.  We consider the case where a given chiral algebra
$\cC$ is a part of a larger algebra $\cA$ that also includes a field
$E(z)$ such that $\oint\! E$ is a screening for $\cC$.  When $E$ is in
addition a fermion, irreducible modules of $\cA$ decompose into entire
complexes of modules over~$\cC$, with the differential
$d\,{=}\,{\oint} E$.  Indecomposable $\cC$-modules~$\modL$ are then
constructed by ``gluing together'' the modules in two adjacent terms
of the complex
$\,\cdots\xrightarrow{d}R_i\xrightarrow{d}R_{i+1}\xrightarrow{d}
\cdots$, in such a way that~$R_i$ becomes a submodule of~$\modL$,
while the quotient~$\modL/R_i$ is isomorphic to~$R_{i+1}$.  The
pre-image of $R_{i+1}$ in the indecomposable module can be thought of
as descendants of a field involving the operator $\dd^{-1}\!E$ by
which the chiral algebra is extended.  Not surprisingly, extending the
chiral algebra by an operator of the form $\dd^{-1}\!E$ leads to the
occurrence of logarithms in operator products.  We apply this
construction of indecomposable modules to $(2,p)$ Virasoro minimal
models and to $\hSL2$ in a realization that admits a fermionic
screening.  In the associated Virasoro representation given by the
Sugawara construction, the generator $L_0$ then acquires dimension-$2$
Jordan cells, which can again be restated as the existence of
``logarithmic partners'' to certain states.\footnote{~The occurrence
  of Jordan cells as the characteristic property of indecomposable
  representations pertains to the Virasoro algebra.  The existence of
  Jordan cells in a Virasoro module associated with an affine Lie
  algebra via the Sugawara construction implies that the affine Lie
  algebra module is indecomposable.}

In its simplest form, our approach involves a deformation of the
energy-momentum tensor that eventually leads to the construction of
indecomposable representations.  This deformation is described in
Sec.~\ref{sec:log-emt}.  A deformation of general chiral algebras is
considered in Sec.~\ref{sec:deform-general}.  In
Sec.~\ref{sec:constr-deform}, we first recall some basic facts about
operator product algebras (OPAs).  Then we proceed to construct an OPA
derivation and use it to perform the deformation (see
Theorem~\ref{thm:deform}).  Thinking of this (outer) derivation as an
inner one generated by a first-order pole with some new operator then
leads to an extension of the chiral algebra.  In Sec.~\ref{sec:K}, we
show how the corresponding space of states is extended such that the
operator--state correspondence is maintained.  The resulting space of
states carries an indecomposable representation of the chiral algebra.
The construction is reformulated and generalized in terms of vertex
operator algebras.  Ward identities for the energy-momentum tensor in
the logarithmic theories constructed via a fermionic screening are
presented in Sec.~\ref{sec:Ward}. Specific examples of the general
construction are studied in Secs.~\ref{sec:examples} (Virasoro models)
and~\ref{sec:log-sl2} (the $\hSL2$ WZW theory). Finally, in
Sec.~\ref{sec:conclusions} we briefly discuss how our results fit into
the quest of gaining a deeper understanding of logarithmic~CFT.

\section{A deformation of the energy-momentum tensor}
\label{sec:log-emt}
As already mentioned, logarithmic \cfts\ are naturally associated with
the presence of Jordan cells in Virasoro modules.  In this section, we
present a simple recipe for deforming the energy-momentum tensor that
leads to the appearance of Jordan cells.

To start, we consider the simple problem of finding an energy-momentum
tensor of the form
\begin{gather}
  \tilde{T}(z)=T(z)+T_1(z) \,,
\end{gather}
where $T$ is the energy-momentum tensor of a (standard) conformal
field theory and $T_1$ is an ``improvement'' term that eventually
leads to Jordan cells. Later on, we construct indecomposable Virasoro
representations by allowing $T_1$ to act on an auxiliary vector space,
but at the present stage, we first ensure that $\tilde{T}$ has the
correct operator product to be an energy-momentum tensor.  A simple
ansatz is to take $T_1$ to have regular OPE with itself and to be a
primary field of some weight~$h$ with respect to $T$.  In this case we
have
\begin{gather}
  \tilde{T}(z)\,\tilde{T}(w)=(T(z)+T_1(z))\,(T(w)+T_1(w))=
  \frac{2\,T(w)+2h\,T_1(w)}{(z-w)^2}+\ldots\,,
\end{gather}
which implies $h\,{=}\,1$.  To also have compatibility between the
grading of Laurent modes, we therefore take $T_1$ to be of the form
\begin{equation}\label{T-through-E}
  T_1(z)=\tfrac 1z\, E(z)\,,
\end{equation}
with $E(z)$ a weight-one primary field.  More general solutions, for
which $T_1(z)\,T_1(w)$ does possess a singular part are also possible.
We further analyze the general case in Sec.~\ref{sec:deform-general}.
Here we continue our construction with the particular deformation
\eqref{T-through-E} for the particular case of a single free boson
with a background charge.  We show that ansatz~\eqref{T-through-E}
leads to selecting the $(1,n)$ and $(2,n)$ minimal models and that it
in fact offers a construction of their logarithmic extension.

\smallskip

Let $\varphi$ be a canonically normalized free boson,
$\varphi(z)\,\varphi(w)\,{=}\,\log(z{-}w)$. Then the energy-momentum
tensor
\begin{equation}
  T(z) =
  \thalf\,{\boldsymbol{:}}{\p\varphi(z)\,\p\varphi(z)}{\boldsymbol{:}}
  + \alph\,\p^2\varphi(z)  \label{3b}
\end{equation}
has the central charge $c\,{=}\,1\,{-}\,12\alph^2$.  The conformal
weight of the vertex operator
${\boldsymbol{:}}\rme^{\gamma\vfi(z)}{\boldsymbol{:}}$ in this model is
(the colon indicating normal ordering is suppressed in what follows)
\begin{equation}
  h(\rme^{\gamma\vfi}) = \thalf\,\gamma^2 - \alph\gamma \,.
  \label{delta}
\end{equation}

We now specialize ansatz~\eqref{T-through-E} for $T_1(z)$ as
\begin{equation}
  T_1(z)=\tfrac1z\, \beta\, \rme^{\gamma \vfi} \,.
  \label{ansatz}
\end{equation}
Here $\gamma$ is a number, while we keep the precise meaning of the
quantity $\beta$ as yet unspecified.  It may be a constant or an
operator in a \findim\ vector space, in which case it accounts for the
presence of the latter auxiliary space with which the \rep\ spaces of
the original theory are to be tensored ($\beta$ can then be regarded
as a kind of zero mode operator, similar to the center of mass
coordinate of a free boson or to the gamma matrices in the Ramond
sector of fermions).  We now determine the most general value of
$\gamma$ for this ansatz to work.  For $T_1$ to have a regular OPE
with itself, i.e.,\footnote{~As usual, $A(z)B(w)$ in the left-hand
  side of operator products stands for the singular part of the
  radially ordered operator product of $A(z)$ and $B(w)$.}
\begin{gather}\label{t2t2}
  T_1(z)\,T_1(w)=\, \frac{1}{zw}\,\beta^2\,(z-w)^{\gamma^2}
  \rme^{\gamma(\varphi(z) + \varphi(w))} = 0 \,, 
\end{gather}
we must require that $\gamma^2\,{=}\,p \in\{0,1,2,...\}$ or that
$\beta^2\,{=}\,0$.

We begin with the first possibility and take $\gamma\,{=}\,\sqrt p$.
The requirement that $\rme^{\gamma \vfi}$ is a weight-one operator
then yields
\begin{equation} \label{sol}
  \alph=\frac{\sqrt p}{2}-\frac{1}{\sqrt p}\,,
\end{equation}
and hence $c\,{=}\,13-3p-\frac{12}p$.  Obviously, for $T_1$ to be
bosonic, $\beta$ must be bosonic (fermionic) if and only if
$\rme^{\gamma \vfi}$ is bosonic (fermionic).  We must therefore
distinguish two cases:
\begin{enumerate}
\item $\beta$ is bosonic and $p\,{=}\,2q$ with $q\,{=}\,1,2,3,...\,$.
  Then (\ref{sol}) gives the $(1,q)$ minimal series
  \begin{equation}
    \alph = \frac{1}{\sqrt{2}}\,(\sqrt q - \frac{1}{\sqrt q})
  \end{equation}
  with the central charge
  \begin{equation}\label{3c}
    c = 13 - 6q -  \tfrac 6q \,.  
  \end{equation}
  In this case, $\beta$ can in particular be just a constant.
  
\item\label{case2} $\beta$ is fermionic and $p\,{=}\,2r{+}1$ with
  $r\,{=}\,0,1,2,...\,$.  This gives the $(2,p)$ minimal series with
  the central charge
  \begin{equation}
    c=10-6r- \tfrac{12}{2r+1} \,.
  \end{equation}
\end{enumerate}
  
\noindent
Note that the value $c\,{=}\,{-}2$ is contained in both series.  In
the first series it corresponds to using the vertex operator with
$\gamma^2\,{=}\,4$ (``long screening''), while in the second case one
has $\gamma^2\,{=}\,1$ (``short screening'').

\medskip

The second option that follows from~\eqref{t2t2} is $\beta^2\,{=}\,0$.
The situation where $\beta$ is a nilpotent boson does not lead to any
restrictions on the central charge, and we do not consider it here. A
more natural case is that of a fermionic~$\beta$, which brings us back
to the above case~\ref{case2}.

To summarize, ansatz~\eqref{ansatz} for $T(z)$ leads to two possible
series of solutions: the $(1,q)$ minimal series, and the $(2,p)$
series with $p$ odd.

The idea on how this deformation gives rise to a logarithmic extension
is to allow $\beta$ to be an operator acting on a suitable auxiliary
space $K$ such that $\tilde{T}$ furnishes an indecomposable
representation of the Virasoro algebra.  The construction of the state
space then depends on the representation chosen for~$\beta$.  In
particular, for $\beta\,{\in}\,\End K$ with $\dim K\,{>}\,2$,
generically we must deal with Jordan cells that can have size larger
than two. (Correspondingly, higher powers of logarithms occur in the
operator products.) In contrast, in the fermionic case we only
encounter Jordan cells of size~two.


\section{Deformation of a general chiral algebra}
\label{sec:deform-general}
In the previous section, we constructed a particular deformation of a
specific conformal field\,---\,the energy-momentum tensor $T$\,---\,in
a special class of models.  We now extend the construction beyond the
one-boson case, not relying on the presence of a free-field
realization, and deforming all fields in the chiral algebra.  This is
done in such a way that the OPEs are preserved, but $\beta$ is
interwoven in the deformed operators such that new, indecomposable
representations are obtained by tensoring with an auxiliary vector
space $K$ and taking $\beta\,{\in}\,\End K$.  The two basic
ingredients of the construction --- the deformation and the extension
of the space of states --- are described in
Secs.~\ref{sec:constr-deform} and~\ref{sec:K}, respectively.

\subsection{The OPA deformation} 
\label{sec:constr-deform}
Instead of working with operator product expansions of the form
\begin{equation}\label{ABope}
  A(z)\,B(w)=\sum_{n\geq1}\frac{\opepole[n][A,B](w)}{(z-w)^n}\,,
\end{equation}
we use \textit{operator product algebras} (OPAs) in the form
introduced in \cite{Thi}.  This constitutes a convenient formulation of
the properties of operator products that facilitates explicit
calculations, and can be regarded as an adaptation of (a part of) the
axioms of vertex operator algebras (VOAs) \cite{FRlm,frhl,kac3}.

The essential ingredients of the OPA setting in~\cite{Thi} are as
follows. The operators (or fields) $A$ form a (super)vector space
$\opaV$, which is graded over $\oZ$.  The OPA structure on $\opaV$ is
given by a collection of bilinear operations
\begin{equation}\label{opepole}
  \opepole[n][~,\;]:\quad \opaV\times\opaV\to\opaV\,,\qquad n\in\oZ\,,
\end{equation}
that are compatible with the grading by conformal
weight\footnote{~That is, when $A$ has weight $h_A$ and $B$ weight
  $h_B$, then the field $[A,B]_n$ has weight $h_A+h_B-n$.  The
  existence of the energy-momentum tensor $T\,{\in}\,\opaV$ is
  understood here; we omit the obvious details.} and satisfy the
``associativity'' condition~\cite[Eq.\,(2.3.21)]{Thi}
\begin{equation}\label{right-composite}
  \opepole[q][A,{\opepole[p][B,C]}] =
  (-1)^{A B}\opepole[p][B,{\opepole[q][A, C]}]
  +\sum_{\ell\geq1}\binom{q-1}{\ell-1}
  \opepole[p+q-\ell][{\opepole[\ell][A,B]}, C] \,.
\end{equation}
(The sign factor $(-1)^{AB}$ is equal to $-1$ if $A$ and $B$ are both
fermionic, and to $+1$ otherwise.)  There is also a distinguished
element $\One\,{\in}\,\opaV$ such that
\begin{equation}
  \opepole[n][\One,A] = \delta_{n,0}\,A  
\end{equation}
for all $A\,{\in}\,\opaV$, and an even linear mapping
$\dd:\opaV\,{\to}\,\opaV$ such that the ``commutativity''
property~\cite[Eq.\,(2.3.16)]{Thi}
\begin{equation}\label{axiom-commute}
  \opepole[n][B,A] = (-1)^{AB}\sum_{i\geq n}\frac{(-1)^i}{(i-n)!}\,
  \dd^{i-n}\opepole[i][A,B]
\end{equation}
is satisfied.  For any two operators $A$ and $B$, their OPE is then
reconstructed from the products $\opepole[n][~,\;]$ as
in~\eqref{ABope}.  In other words, operator product~\eqref{ABope}
should be regarded as a linear mapping
\begin{gather}  \label{OtoO}
  \opaV\tensor\opaV \,\to\, \opaV\tensor_{\oC}^{}\oC[[z,z^{-1}]]
\end{gather}
that furnishes a generating function for the infinite family
of products in Eq.~\eqref{opepole}.

To perform the deformation, we fix an operator $E\,{\in}\,\opaV$ and
define the operation
\begin{equation}\label{Delta-E}
  \Delta_E:\quad A(z)\;\mapsto\; \log z\,\opepole[1][E,A]\!(z) +
  \sum_{n\geq1}\frac{(-1)^{n+1}}{n}\,
  \frac{\opepole[n+1][E,A]\!(z)}{z^n} 
\end{equation}
on $A\,{\in}\,\opaV$.

\begin{Thm}\label{thm:deform}
  For any $E\,{\in}\,\opaV$, the operation $\Delta_E$ is a
  superderivation of the OPA $\opaV$.
\end{Thm}
\noindent The proof  is obtained by direct calculation; it is given in
Appendix~\ref{app:proof}.

It follows from Theorem~\ref{thm:deform} that if $E$ is fermionic and
accordingly $\beta$ is also fermionic, so that $\beta^2\,{=}\,0$,
deformation \eqref{Delta-E} gives rise to an OPA isomorphism
\begin{equation}
  A\,\mapsto\, A + \beta\, \Delta_E(A)\,.
\end{equation}
If $E$ is bosonic and $\beta$ is bosonic and nilpotent,
$\beta^N{=}\,0$ for some integer $N$, then
\begin{equation}
  A\,\mapsto\, \exp(\beta\Delta_E) A \,,
\end{equation}
where the right-hand side is a \textit{polynomial} in $\Delta_E$, is
again an isomorphism.  (In a number of specific examples, e.g.\ for
models based on an affine Lie algebra, and with $E$ taken to be a
current in a nilpotent subalgebra, the right-hand side is a polynomial
in $\beta$ without requiring $\beta$ to be nilpotent.)

If $E$ is a weight-$1$ \textit{primary} field (and hence,
$\opepole[n][E,T]{=}\,\delta_{n,2}E$ for $n\,{>}\,0$), the
deformation of the energy-momentum tensor is given by
\begin{equation}\label{T-deform}
  \tilde{T}=\exp(\beta\Delta_E)\,T
  = T + \tfrac{1}{z}\,\beta E +\beta^2\,(\dots)+\dots \,.
\end{equation}
The term linear in $\beta$ reproduces the particular deformation in
Eq.~\eqref{T-through-E} that was considered in the one-boson case in
Sec.~\ref{sec:log-emt}.  In Sec.~\ref{sec:K}, we show that taking
$\beta$ to be an endomorphism of a finite-dimensional space results
in an indecomposable module over the algebra
$\tilde{\cC}\,{:=}\,\exp(\beta\Delta_E)\cC$ that in particular 
contains the Virasoro algebra corresponding to $\tilde{T}$.

In general, the proposed deformation \eqref{Delta-E} maps each field
in the algebra into a linear combination of fields weighted with
powers of~$z$ and~$\log z$.  The product \eqref{OtoO} of the operator
algebra $\opaV$ is thereby lifted to an operator algebra $\talgO$ for
which the operator products furnish a mapping
(cf.~\cite{mila3})\,\footnote{~Since the operation $\Delta_E$ affects
  the Laurent expansion of fields, it drastically changes the operator
  product expansions~\eqref{ABope}. To put it differently, we deal
  with a deformation of the operator product expansion~\eqref{ABope},
  which is however not arbitrary, but such that the products
  $\opepole[n][~,\,]$ in Eq.~\eqref{opepole} are preserved.}
\begin{equation}\label{ope-with-log}
  \talgO\tensor\talgO \,\to\,
  \talgO\tensor_{\oC}^{}\oC[[z,z^{-1}]][\log z]\,.
\end{equation}
The terms involving $\log z$ do not appear in the transformed fields 
if and only if $\opepole[1][E,A]{=}\,0$ for all $A\,{\in}\,\opaV$, 
or in other words, if and only if $\oint\!E$ is a \textit{screening
operator} for the chiral algebra (in that case, although
$\exp\beta\Delta_E$ does not preserve the vector space $\opaV$, it
maps $\opaV$ into an isomorphic OPA $\tilde\opaV$).
Whenever $\oint\!E$ is in addition a fermionic screening, it can serve
as a differential in a complex; the logarithmic deformations of
modules that appear in Sec.~\ref{sec:K} then involve elements of the
adjacent members of the complex.

In applications, screenings are often dealt with as follows.  One
starts by taking some ``large'' algebra, then chooses in it field(s)
whose integrals are declared to be screening(s), and finally selects a
``small'' algebra as the centralizer (commutant) of the screenings in
the large algebra.  Verifying that a given operator $E$ gives a
screening upon integration involves only the first-order poles
$\opepole[1][E,\;\cdot\;]$, while all the order-$n$ poles
$\opepole[n][E,\;\cdot\;]$ with $n\,{\geq}\,2$ can be arbitrary and
are irrelevant to the properties of a conformal field theory in the
commutant of~$\oint\!E$.  However, these poles are ``reanimated'' by
the logarithmic deformation prescription, where their actual role is
to be responsible for ``logarithmic deformation directions'' of the
given conformal field theory.

\subsection{Remarks}
\subsubsection*{a} We can think of the mapping
\eqref{Delta-E} as induced by the operator product of $A$ with some
new field, which we call $\tilde\alpha$. Taking the derivative, we
then see that $\dd \tilde\alpha(z)\,{=}\,{-}E(z)$. We therefore 
use the symbol $\dd^{-1}\!E(z)$ as a suggestive notation for
$-\tilde\alpha(z)$. We can then write
\begin{equation}  \label{Eext}
  \dd^{-1}\!E(z)
  \,A(w)= \opepole[1][E,A]\log(z\,{-}\,w)
  + \sum_{n\geq1}\frac{(-1)^{n+1}}{n}\,
  \frac{\opepole[n+1][E,A]\!(w)}{(w\,{-}\,z)^{n}}\,.
\end{equation}
However, the relation $\dd \tilde\alpha(z)\,{=}\,{-}E(z)$ does not
determine the zero mode of $\tilde\alpha(z)$; as we see in the next
subsection, such zero modes are expressed in terms of endomorphisms of
an auxiliary space.

We note that if $E$ is replaced with $\dd E$, the corresponding
derivation can be written as
\begin{equation*}
  \Delta_{\dd E} A = \opepole[1][{\opepole[0][\cO_1, E]}, A],
\end{equation*}
with $\cO_1(z)\,{=}\,\frac1z$. This operation is obviously a
derivation, because the first-order pole with \textit{any} operator is
a derivation of the OPA.  In contrast, expressing $\Delta_{E} A$ in a
similar way as $\Delta_{E} A\,{=} {\opepole[1][{\opepole[0][\cO_1,
    \dd^{-1}E]}, A]}$ involves an operator not from $\opaV$ whenever
$E\,{\notin}\,\dd\opaV$.


\subsubsection*{b} If $E$ in \eqref{Delta-E} is not of weight~$1$, the
parameter $\beta$ must be dimensionful, and the construction of
$\exp(\beta\Delta_E)$ as a series in $\beta$ involves growing orders
of derivatives and growing powers of $z^{-1}$.  For instance, taking
the Virasoro algebra, which in the OPA language is generated by a
single element $T$ such that
\begin{equation}
  \opepole[1][T,T]=\dd T\,,\quad
  \opepole[2][T,T]=2 T\,,\quad
  \opepole[4][T,T]=\tfrac{c}{2}\,,
\end{equation}
we have
\begin{align}
  \Delta_T T &= \log z\,\dd T + \frac{2}{z}\,T +
  \frac{c}{6z^3}\,,\\
  (\Delta_T)^2 T &= (\log z)^2\, \dd^2 T + 5\, \frac{\log z}{z}\,\dd
  T + \frac{4 - 2\log z}{z^2}\, T +
  c\,\frac{\frac{1}{3} - \half\log z}{z^4}\,,\\
  (\Delta_T)^3 T &= (\log z)^3\,\dd^3 T + 9\,\frac{(\log
    z)^2}{z}\,\dd^2 T
  - \frac{7(\log z)^2 - 19\log z}{z^2}\,\dd T\\
  &\phantom{{}={}} + \frac{4(\log z)^2 - 14\log z + 8}{z^3}\,T +
  c\,\frac{2(\log z)^2 - \frac{17}{6}\log z + \frac{2}{3}}{z^5}\,.
  \notag
\end{align}

  
\subsubsection*{c} As an example that clearly displays the isomorphism
property of $\exp(\beta\Delta_E)$ (although is not relevant to
logarithmic theories), we consider the situation where $E$ is a
bosonic \textit{current}, that is, a weight-$1$ field with the OPEs
\begin{align}\label{current-ope}
  E(z)\,E(w)&=\frac{\varkappa}{(z-w)^2}\,,\\
  T(z)\,E(w)&=\frac{\lambda}{(z-w)^3} + \frac{E(w)}{(z-w)^2} +
  \frac{\dd E(w)}{z-w}
\end{align}
(where the constant $\lambda$ is a possible ``anomaly''), and moreover,
$E$ is diagonal on all primaries of the theory, i.e.,
\begin{equation}
  E(z)\,\Psi_i(w)=\frac{q_i\,\Psi_i(w)}{z-w}\,.
\end{equation}
The operation $\exp(\beta\Delta_E)$ is then readily evaluated to act as
\begin{align}\label{spflow-gen1}
  \exp(\beta\Delta_E)\,E(z)&=E(z)+
  \beta\, \tfrac{\varkappa}{z}\,,\\
  \exp(\beta\Delta_E)\,T(z)&=T(z)+\tfrac{\beta\,E(z)}z
  +\tfrac{\beta\lambda+\beta^2\varkappa}{2z^2}\,,\\
  \exp(\beta\Delta_E)\,\Psi_i(z)&=z^{\beta q_i}\,\Psi_i(z)\,.
  \label{spflow-gen3} 
\end{align}  
(In the last formula, an infinite series of powers of $\log z$ is
summed up to~$z^{\beta q_i}$.)  With $\beta$ being just a number (or,
in the setting of the next subsection, $\beta\in\End(\oC)$), we
recognize this as the spectral flow transform associated with the
current $E$.  

\subsection{Extension of the state space} \label{sec:K}
The observation leading to Eq.~\eqref{Eext} indicates that for
$E\,{\notin}\,\dd\opaV$, the deformation can be described in terms of
a new field introduced into the OPA such that the outer derivation
$\Delta_E$ becomes an inner one.  We therefore consider an extended
OPA containing this new field, assuming a definite prescription for
the zero mode.  In order to preserve the field--state correspondence,
we must then also extend the space of states (the vacuum module).

The case considered in~\eqref{T-deform} already captures crucial
features of the general situation. We take the vacuum module $V$ of
the OPA $\opaV$, with the highest-weight vector $\Vac_V$, and extend
it by taking the tensor product with some auxiliary finite-dimensional
vector space $K$.  As a module over $\opaV$, $V\Tensor K$ is just the
direct sum of ${\rm dim}\,K$ copies of $V$, with operators in
$\End(K)$ generically acting between different copies. For a chosen
$\vac\,{\in}\,K$, we can then identify $\Vac_V$ with the vector
$\Vac=\Vac_V\Tensor\vac$.  In the deformation formulas, we now take
$\beta$ to be an element of~$\End(K)$ such that
\begin{gather}\label{alpha-kill}
  \beta\,\Vac\,{=}\,0 \,.
\end{gather}
We give the operators in $\End(K)$ the same parity (bosonic or
fermionic) as $E$ (so that in particular $\beta E$ is always bosonic).
Now, let $\alpha\,{\in}\,\End(K)$ satisfy
\begin{equation}\label{alpha-beta-1}
  [\alpha,\beta]= 1
\end{equation}
with $[\;,\,]$ denoting the \textit{super}commutator.
{}From~\eqref{T-deform} (where $E$ is a weight-$1$ field), the modes
$\tilde{L}_n$ of the deformed energy-momentum tensor $\tilde{T}$ act
on $\alpha\,\Vac$ as
\begin{equation}
  \tilde{L}_n\cdot\alpha\,\Vac=
  (T + \tfrac{1}{z}\,\beta E)_n\cdot\alpha\,\Vac\,.
\end{equation}
In particular,
\begin{equation}
  \tilde{L}_{-1}\cdot\alpha\,\Vac= -E_{-1}\Vac\,.
\end{equation}
Via the field--state correspondence, the state in the right-hand side
is associated to the field $E(z)$.  We therefore conclude that
\textit{the operator $\tilde\alpha(z)$ corresponding to the state
$\alpha\,\Vac$ satisfies}
\begin{equation}\label{dA=E}
  \dd\tilde\alpha(z)= -E(z) \,,
\end{equation}
which we also write suggestively as 
$\tilde\alpha(z)\,{=}\,{-}\dd^{-1}\!E(z)$.

This illustrates how the space of states (and hence, the fields) can
be extended.  We now analyze the construction systematically, still
assuming $\beta$ to be an endomorphism of an auxiliary
finite-dimensional vector space~$K$ (with $h_E\,{=}\,1$).  The
necessary manipulations are best described in the language of vertex
operator algebras~\cite{FRlm,frhl,kac3}, which we summarize in
Appendix~\ref{voa}.

We begin with making the auxiliary vector space $K$ into a ``toy''
vertex operator algebra. Having done so allows us to use the
products $\opepole[n][~,\,]$ of the corresponding ``fields,'' see
Eqs.~\eqref{bracket-new} and~\eqref{bracket-neg}.  These products on
$\End(K)\aff$, satisfying all the properties of the OPA operations
except the ``commutativity'' condition, then allow us to construct a
generalization of $\Delta_E$ in~\eqref{Delta-E}.

We first choose an arbitrary element $\vac\,{\in}\, K$, which we call
the vacuum vector of $K$.  This gives rise to a vector space
decomposition $\End K\,{=}\,(\End K)_+ + (\End K)_-$, where $(\End
K)_+$ is defined by the requirement that $(\End K)_+\vac\,{=}\,0$. We
next fix a set $\cS$ of elements from $(\End K)_-$ such that every
vector in $K$ can be written as a linear combination of vectors
$k\,{=}\,\kappa\vac$ with $\kappa\,{\in \cS}\,$.  Elements of $(\End
K)_+$ are said to be annihilation operators and elements of $\cS$
creation operators. {}From the (super)commutator in $\End(K)$ and the
normal ordering associated with the chosen vacuum vector, we can then
define an \textit{OPA structure on} $\End(K)\aff$ by
Eqs.~\eqref{bracket-new}\,--\,\eqref{bracket-0}.

Moreover, setting $Y(k,z)\,{=}\,\kappa$ for $k\,{=}\,\kappa\vac$ with
$\kappa\,{\in}\,\cS$ defines a (toy) vertex operator algebra~$\voaK$
with the space of states given by~$K$.  The
data~$\voaK\,{=}\,(K,Y,\vac)$ satisfy the list of properties of VOAs
except those related to the Virasoro algebra.  Here, $z$ is a dummy
variable because $Y$ maps $K$ into $\End(K)$ and the corresponding OPA
$\opaK$ actually coincides with $(\End K)_-$ (with the only nonzero
bracket being just $\opepole[0][A,B]$ given by the multiplication
in~$(\End K)_-$).

Returning to the ``genuine'' VOAs, we fix a VOA $\voaV'$ with the
space of states~$V'$ and the corresponding OPA~$\opaV'$. Selecting a
weight-$1$ field $E(z)\,{\in}\,\opaV'$ (in general, several such
fields), we define a VOA $\voaV\,{\subset}\,\voaV'$ as the commutant
(centralizer) of its zero mode (in general, of the subalgebra of zero
modes); we thus have ~$\opepole[1][E,A]{=}\,0$ for every
field~$A\,{\in}\,\opaV$. For a VOA module $\modW'$, we similarly
select a submodule~$\modW\, {\subset}\,\modW'$.  Clearly, $\voaV'$ is
a module over~$\voaV$.

Our strategy is now to construct a suitable deformation and extension
of the $\voaV$-modules~$\voaV'$ and~$\modW'$ by taking the tensor
products~$\voaV'\Tensor\voaK$ and~$\modW'\Tensor\voaK$, where the role
of the auxiliary sector $\voaK$ is to introduce $\log z$ into the
theory. To achieve this goal, we replace $\End(K)\aff$ 
with the space $\End(K)\aff[\log z]$ and choose an element
$\varepsilon\,{\in}\,\End(K)\aff[\log z]$ of the form
\begin{equation}\label{epsilon}
  \varepsilon=\beta\log z\,{+}\,v \,,
  \qquad\beta\,{\in}\,\End K\,,
  \quad v\,{\in}\,\End(K)\aff
\end{equation}
such that the limit $\lim_{z\to0}\varepsilon\,\vac$ exists. (This
condition requires in particular that $\beta\vac\,{=}\,0$, which
corresponds to~\eqref{alpha-kill}. Note that $z$ is really a formal
variable here, i.e., the existence of the limit implies that
$\varepsilon\vac\,{\in}\,K[[z]]$. Also, $\log z$ is to be regarded as
\textit{another} formal variable that has the additional property that
$\dd(\log z)\,{=}\,z^{-1}$~\cite{mila3}.)

Then for~$U\,{=}\,V'$ or~$W'$, we define the mapping
\begin{equation}\label{eq:the-derivation}
  \begin{split}
    \Delta_{E,\varepsilon}:\quad
    \Hom&(V',U)\aff{\tensor}_{\oC\aff}^{\phantom{y}}
    \End(K)\aff[\log z]\\
    {}&\longrightarrow\,\Hom(V',U)\aff{\tensor}_{\oC\aff}^{\phantom{y}}
    \End(K)\aff[\log z]\\
    A\; &\longmapsto\,
    \sum_{n\geq0} \opepole[-n][\varepsilon,{\opepole[n+1][E,A]}]
    + (-1)^{E} \sum_{n\geq0}
    \opepole[-n][\dd^{-1}\!E,{\opepole[n+1][\dd \varepsilon,A]}]\,,
  \end{split}
\end{equation}
which depends on the chosen field (``screening current'')
$E\,{\in}\,\opaV'$ and on $\varepsilon$ of the form~\eqref{epsilon}.
All the products $\opepole[n][~,\,]$ are well-defined with the help of
Eqs.~\eqref{bracket-new}--\eqref{bracket-0}.
\begin{Lemma}\label{thm:general}
  The mapping $\Delta_{E,\varepsilon}$ defined in
  \eqref{eq:the-derivation} with $\varepsilon$ of the form
  \eqref{epsilon} satisfies the following properties:
  \begin{enumerate}

  \item\label{item:NEW}
    It generalizes the mapping~\eqref{Delta-E}: for $A\,{\in}\,\opaV'$
    or $\opaW'$ we have
    \begin{equation}\label{eq:NEW}
      \Delta_{E,\beta\log z}A = \sum_{n\geq0}
      \frac{\beta}{n!}\,(\dd^n\log z)\, \opepole[n+1][E,A],
    \end{equation}
    and therefore, $\Delta_{E,\beta\log z}\,{=}\,\beta\Delta_E$ on
    elements of $\opaV'$ or $\opaW'$.
    
  \item\label{item:2} For $A\,{\in}\,\opaV$,
    \begin{equation}\label{Delta-reduce}
      \Delta_{E,\varepsilon}A=\sum_{n\geq1}
      \opepole[-n][\varepsilon,{\opepole[n+1][E,A]}],
    \end{equation}
    and moreover, $\Delta_{E,\varepsilon}$ is a derivation of the OPA
    subalgebra $\opaV$,\footnote{~Clearly, the
      OPA~$\opaV\simeq\opaV\Tensor\id_K$ is a subalgebra
      in~$\opaV'\Tensor\End(K)\aff[\log z]$.}
    \begin{equation*}
      \Delta_{E,\varepsilon}\opepole[n][A,B]=
      \opepole[n][{\Delta_{E,\varepsilon}A},B] +
      \opepole[n][A, {\Delta_{E,\varepsilon}B}]\qquad{\rm for}\quad
      A,B\,{\in}\,\opaV\,.
    \end{equation*}
    
  \item\label{item:3} The OPA
    $\;\tilde\opaV\,{:=}\,\exp\Delta_{E,\varepsilon}\opaV$ is
    isomorphic to~$\opaV$.
    

  \end{enumerate}
\end{Lemma}

To prove assertion~\ref{item:NEW}, we note that for $A\,{\in}\,\opaV'$
or $\opaW'$, each term in the second sum in~\eqref{eq:the-derivation}
vanishes when applied to elements of the form~$A\,{\tensor}\one$
with~$A\,{\in}\,\opaV'$, while in accordance with~\eqref{bracket-neg},
the first sum can be rewritten as in Eq.~\eqref{eq:NEW}.  When we
further specialize to elements of the form~$A\,{\tensor}\one$
with~$A\,{\in}\,\opaV$, the term corresponding to $n\,{=}\,0$ in the
first sum also vanishes because~$E$ is a screening current for~$\opaV$.
This then shows~\eqref{Delta-reduce} for $A\,{\in}\,\opaV$.
The proof that $\Delta_{E,\varepsilon}$ is a derivation of~$\opaV$ is
a mere reformulation of the proof of Theorem~\ref{thm:deform} given in
Appendix~\ref{app:proof} (using~\eqref{bracket-neg} with
$\cO_0(z)\,{=}\,\beta\log z$).  This also shows that the OPA
$\tilde\opaV\,{=}\,\exp\Delta_{E,\varepsilon}\opaV$ is isomorphic
to~$\opaV$.\hfill\rule{1.5ex}{1.5ex}

We also remark that the VOA $\tilde\voaV$ that corresponds to the OPA
$\tilde\opaV$ is isomorphic to~$\voaV$. Moreover, the vector spaces
$\tilde\opaV'\,{:=}\,\exp\Delta_{E,\varepsilon}\!\left(\opaV'{\tensor}
_{\oC\aff}\opaK\right)$ and $\tilde\opaW'\,{:=}\,\exp\Delta_
{E,\varepsilon}\!\left(\opaW'{\tensor}_{\oC\aff}\opaK\right)$ are 
modules over the OPA
$\tilde\opaV$, while the corresponding VOA~$\tilde\voaV'$ and
similarly $\tilde\modW'$ are modules over the VOA~$\tilde\voaV$. As
$\tilde\voaV$-modules, $\tilde\voaV'$ and~$\tilde\modW'$ contain
submodules that are isomorphic to~$\voaV'$ and~$\modW'$, respectively.
This follows from the fact that~$\opaV'\,{\subset}\, \tilde\opaV'$
and~$\opaW'\,{\subset}\,\tilde\opaW'$ can be identified with elements
of the form~$A\tensor\vac$, where $A\,{\in}\,\opaV'$ or~$\opaW'$, with
the derivation~$\Delta_{E,\varepsilon}$ taking the
form
\begin{equation*}
  \Delta_{E,\varepsilon}A=\sum_{n\geq0}
  \opepole[-n][\varepsilon,{\opepole[n+1][E,A]}]
\end{equation*}
on these subspaces.

\medskip

The applications considered in this paper pertain to the case where an
element $\alpha$ is chosen in $\End K$ such that
\begin{align}
  \opepole[1][{\dd\varepsilon},\alpha] &= -(-1)^E\id_K\,,
  \label{de-alpha}\\
  \opepole[n][{\dd\varepsilon},\alpha] &= 0\,,\qquad n\,{\geq}\,2 \,.
\end{align}
(For $\varepsilon$ of the form~\eqref{epsilon}, it follows from the
identity~\eqref{bracket-new} that Eq.~\eqref{de-alpha} is equivalent 
to $[\beta,\alpha]\,{=}\,{-}(-1)^E\id_K$.) We then define the field
\begin{equation}
  \tilde\alpha(z) := \rme^{\Delta_{E,\varepsilon}}\alpha\,.
\end{equation}
With~$E$ assumed, for simplicity, to be OPE-isotropic, i.e.,
$\opepole[n][E,E]{=}\,0$ for $n\,{\geq}\,1$, we find
\begin{equation}
  \tilde\alpha(z) = \alpha - \dd^{-1}\!E(z)\tensor\id_K
  \equiv \alpha - \dd^{-1}\!E(z)\,,
\end{equation}
which in particular leads to~\eqref{dA=E}.
In Sec.~\ref{sec:examples}, we consider examples where the operator
$\tilde\alpha(z)\,{=}\,{-}\dd^{-1}\! E(z)$ naturally arises in this way.

\subsection{Ward identities}
\label{sec:Ward}
We now derive Ward identities for the energy-momentum tensor in
logarithmic theories obtained via the above deformation, in the case
where~$\oint\!E$ is a fermionic screening, $K\,{=}\,\oC^2$,
and~$\End(\oC^2)$ is generated by fermionic operators~$\alpha$
and~$\beta$ satisfying~$\{\alpha,\beta\}\,{=}\,1$.

We consider a VOA $\voaV$ whose space of states~$V$ is a vacuum
representation of the Virasoro algebra and fix a set of
modules~$\modW_i$ over~$\voaV$. Let~$W_i$ be the corresponding
representation spaces and
$W\,{=}\,V\,{\oplus}\,W_1\,{\oplus}\cdots{\oplus}\,W_n$ be the (R)CFT
space of states. In the models of our main interest, $V$ and~$W_i$ are
defined as the cohomology of the screening~$\oint\!E$ acting in larger
spaces~$V'$ and~$W'_i$; the basic case where these structures can be
realized is with the corresponding VOA~$\voaV'$ being a free-field
realization of the Virasoro algebra, but it can also be some other
algebra.  As we have seen, choosing~$\varepsilon\,{=}\,\beta\log z$
results in extending the operators by an element $\tilde\alpha$ such
that~$\dd\tilde\alpha\,{=}\,{-}E$.

Let $\Psi_i$ be primary fields corresponding to the spaces $W_i$,
which satisfy
\begin{equation}
  T(z)\,\Psi_i(w)=\frac{h_i\Psi_i(w)}{(z-w)^2}+
  \frac{\dd\Psi_i}{z-w}\,.
\end{equation}
The deformed theory also involves the fields
\begin{equation}
  \tilde\Phi_i=-{\boldsymbol{:}}\tilde\alpha\tilde\Psi_i{\boldsymbol{:}}
\end{equation}
(where~$\tilde A\,{\equiv}\,\rme^{\Delta_{E,\varepsilon}}\!A$ for every
field~$A$) whose OPEs with the (deformed) energy-momentum tensor are
given by
\begin{gather}
  \tilde T(z)\,\tilde\Phi_i(w)=
  \sum_{m\geq3}\frac{[\tilde E,\tilde\Psi_i]_{m-1}}{(z-w)^m}
  +\frac{h_i\tilde\Phi_i+[\tilde E,\tilde\Psi_i]_1}{(z-w)^2}
  +\frac{\dd\tilde\Phi_i}{z-w}\,.
\end{gather}

This shows the appearance of Jordan cells spanned by $\tilde\Phi_i$ and
$[\tilde E,\tilde\Psi_i]_1$.  The field $[\tilde E,\tilde\Psi_i]_1$ is
not primary in general.  However, such fields seem to provide a
natural extension of the concept of primary fields to logarithmic
models.  We have
\begin{equation}
  \tilde T(z)\,[\tilde E,\tilde\Psi_i]_n(w)=
  -\sum_{m\geq3}\frac{(n{-}1)[\tilde E,\tilde\Psi_i]_{m+n-2}}{(z-w)^m}
  +\frac{(h_i{-}n{+}1)[\tilde E,\tilde\Psi_i]_n}{(z-w)^2}
  +\frac{\dd[\tilde E,\tilde\Psi_i]_n}{z-w}\,.
\end{equation}

Converting the above OPEs into differential equations (Ward
identities) for correlation functions requires introducing a scalar
product on the space of states~$W'\,{=}\,V'{\oplus}
W'_1{\oplus}\cdots{\oplus} W'_N$.  With $\Vac$ denoting the vacuum
vector in~$V'$, we assume that as a part of the definition of the
corresponding conformal field theory, the space~$W'$ is equipped with
a scalar product~$(\,\cdot\,,\,\cdot\,)'$ that is invariant with
respect to the VOA~$\voaV'$ and satisfies~$(\Vac\,,\,\Vac)'\,{=}\,1$.
The product~$(\,\cdot\,,\,\cdot\,)'$ determines a Hermitian
conjugation~${}^{\dagger'}$, which is an involutive anti-automorphism
of~$\voaV'$.

The above product is a natural scalar product for the ``large'' VOA.
We now consider the scalar product $(\,\cdot\,,\,\cdot\,)$ on~$W'$
that is naturally associated to a ``smaller'' (e.g., free-field)
VOA~$\voaV$ (Virasoro in the situation considered here) in the sense
that it is invariant with respect to~$\voaV$ and that the
corresponding conjugation~${}^\dagger$ is an involutive
anti-automorphism of~$\voaV$.  For the Virasoro algebra, this implies
the well-known conjugation relation
\begin{gather}
  T(z)^\dagger=\tfrac{1}{z^4}\,T(\tfrac{1}{z})\,,
\end{gather}
while for the chosen element $E$ we impose
\begin{equation}
  E(z)^\dagger=\tfrac{1}{z^2}\,E(\tfrac{1}{z})\,.
\end{equation}

The correlation functions in the models corresponding to~$\voaV$
and~$\voaV'$ are then defined as
\begin{equation}\label{corfu}
  \bigl<A_1(z_1)A_2(z_2)\cdots A_K(z_K)\bigr>
  =(\Vac,A_1(z_1)A_2(z_2)\cdots A_K(z_K)\Vac)
\end{equation}
and
\begin{equation}\label{corfu'}
  \bigl<A_1(z_1)A_2(z_2)\cdots A_K(z_K)\bigr>'
  =(\Vac,A_1(z_1)A_2(z_2)\cdots A_K(z_K)\Vac)'
\end{equation}
respectively. It is well known that these correlation functions are
related as follows. We let $\ketinfty$ denote the state that
satisfies\,\footnote{~Of course, $\ketinfty$ is not uniquely determined
  by~\eqref{ketinfty}, but in specific models there is usually a
  preferred choice.}
\begin{equation}\label{ketinfty}
  (\ketinfty\,,\Vac)=1\,.
\end{equation}
Then in accordance with the field--state correspondence, there exists
a field~$\Psi(z)$ such that
$\ketinfty\,{=}\,\lim_{z\to0}\!\Psi(z)\Vac$, and correlation
functions~\eqref{corfu} can be rewritten in terms of those
in~\eqref{corfu'}~as
\begin{multline}
  \bigl<A_1(z_1)A_2(z_2)\cdots A_K(z_K)\bigr>
  =(\ketinfty,A_1(z_1)A_2(z_2)\cdots A_K(z_K)\Vac)'=\\
  =\bigl<\Psi(\infty)A_1(z_1)A_2(z_2)\cdots A_K(z_K)\bigr>'.\quad
\end{multline}

In the module $\tilde W'$, i.e., the space of states of the
VOA~$\tilde\voaV'$ introduced after the proof of
Lemma~\ref{thm:general}, we again have two distinct scalar products.
Abusing notation, we still denote one of them
by~$(\,\cdot\,,\,\cdot\,)'$; this is simply the scalar product
on~$W'{\tensor}\,\oC^2_{}$, where~$\oC^2$ is endowed with the scalar
product such that
\begin{equation}
  (\vac,\vac)'=1\qquad{\rm and}\qquad \alpha^{\dagger'}=\beta
\end{equation}
(recall that the vacuum in $\oC^2$ obeys $\beta\vac\,{=}\,0$ and
$\alpha\vac\,{\neq}\,0$). We then
have~$(\Vac\Tensor\vac,\Vac\Tensor\vac)'\,{=}\,1$.

The scalar product $(\,\cdot\,,\,\cdot\,)_L$ of the logarithmic model
is now defined by the relations
\begin{gather}
  (\alpha\ketinfty,\Vac)_L=1 \qquad{\rm and}\qquad (\tilde 
  A^{\dagger_L}\Ket{x}\,,\,\Ket{y})_L=(\Ket{x}\,,\tilde A\Ket{y})_L\,.
\end{gather}
By construction, the ``logarithmic'' scalar
product~$(\,\cdot\,,\,\cdot\,)_L$ is invariant with respect to the
VOA~$\tilde\voaV$; the corresponding Hermitian
conjugation~${}^{\dagger_L}$ is therefore an involutive
anti-automor\-phism of~$\tilde\voaV$.  In the particular case of the
Virasoro algebra, this requires
\begin{gather}
  \alpha^{\dagger_L}=-\alpha\,,\qquad\beta^{\dagger_L}=-\beta
\end{gather}
and~${}^{\dagger_L}={}^\dagger$ on $\voaV'$. This in turn implies that
\begin{equation}
  \tilde\alpha(z)^{\dagger_L}=-\tilde\alpha(\tfrac{1}{z})\,,
\end{equation}
which is consistent with the above ${}^\dagger$-conjugation of $E(z)$
and Eq.~\eqref{dA=E}.  

We are now in a position to consider correlation functions
\begin{equation}
  \bigl<\tilde A_1(z_1)\tilde A_2(z_2)\cdots\tilde A_K(z_K)\bigr>_{\!L}
  =(\Vac,\tilde A_1(z_1)\tilde A_2(z_2)\cdots\tilde A_K(z_K)\Vac)_L
\end{equation}
of the logarithmically extended theory. These can also be rewritten as
\begin{multline}
  \bigl<\tilde A_1(z_1)\tilde A_2(z_2)\cdots \tilde A_K(z_K)\bigr>_L
  =(\alpha\ketinfty,
  \tilde A_1(z_1)\tilde A_2(z_2)\cdots\tilde A_K(z_K)\Vac)'\\
  =\bigl<{\boldsymbol{:}}\tilde\alpha
  \tilde\Psi{\boldsymbol{:}}(\infty)\,\tilde A_1(z_1)\tilde A_2(z_2)
  \cdots \tilde A_K(z_K)\bigr>'
\end{multline}
in terms of the $\langle\cdots\rangle'$ correlation functions.

For the fields $\tilde{\Psi}_i$ and $\tilde\Phi_i$ that span Jordan 
cells with respect to $\tilde{L}_0$, we then have the Ward identities
\begin{multline}
  \bigl<\tilde T(z)\tilde\Psi_I(z_{I})\tilde\Psi_{I+1}(z_{I+1})
  \cdots\tilde\Psi_{I+K}(z_{I+K})\tilde\Phi_{J}(u_{J}) \tilde\Phi
  _{J+1}(u_{J+1}) \cdots \tilde\Phi_{J+L}(u_{J+L})\bigr>_{\!L}\\
  \shoveleft{=\Bigl(
    \sum_{i=I}^{I+K}\bigl(\frac{\Delta_i}{(z-z_i)^2}+\frac{1}{z-z_i}
    \frac{\dd}{\dd z_i}\bigr)+
    \sum_{i=J}^{J+L}\bigl(\frac{\Delta_i}{(z-u_i)^2}+\frac{1}{z-u_i}
    \frac{\dd}{\dd u_i}\bigr)\Bigr)}\times{}\\
  \shoveright{
    {}\times\bigl<\tilde\Psi_I(z_{I})\tilde\Psi_{I+1}(z_{I+1})
    \cdots\tilde\Psi_{I+K}(z_{I+K})\tilde\Phi_{J}(u_{J})
    \tilde\Phi_{J+1}(u_{J+1})\cdots\tilde\Phi_{J+L}(u_{J+L})\bigr>_1}\\
  \shoveleft{\quad{}+ \sum_{i=J}^{J+L}\sum_{n\geq2}\frac{1}{(z-u_i)^n}
    \,\bigl<\tilde\Psi_I(z_{I})\tilde\Psi_{I+1}(z_{I+1})
    \cdots\tilde\Psi_{I+K}(z_{I+K})}\\[-5pt]
  {}\times\tilde\Phi_{J}(u_{J})\tilde\Phi_{J+1}(u_{J+1})\cdots
  [\tilde E, \tilde\Psi_i]_{n-1}(u_i)\cdots
  \tilde\Phi_{J+L}(u_{J+L})\bigr>_{\!L}\,.\;\
\end{multline}

\section{Logarithmic Deformations: $(2,p)$ Virasoro Examples}
\label{sec:examples}
In this section, we apply the above strategy of constructing
logarithmic representations of conformal field theories to Virasoro
minimal models that admit a fermionic screening.  Adding the screening
current to the algebra then results in a larger algebra whose modules
contain the entire Felder complex~\cite{feld} of Virasoro modules.
The zero mode of one of the extended algebra generators is therefore
the differential (BRST operator) of the Felder complex of Virasoro
modules. This picture admits a nice extension to logarithmically
deformed theories.

\subsection{The $c\,{=}\,{-}2$ theory: From free fermions to symplectic
  fermions} \label{sec:c=-2}
We take the Virasoro algebra with central charge $c\,{=}\,{-}2$ and
extend it by the fermionic screening current to the model of free
fermions $\xi$ and $\eta$ with the operator product
\begin{equation} \label{xi-eta}
  \xi(z)\,\eta(w)=\frac{1}{z-w}\,
\end{equation}
(and $\xi(z)\,\xi(w)\,{=}\,0\,{=}\,\eta(z)\,\eta(w)$).  The
$c\,{=}\,{-}2$ Virasoro algebra is then generated by the modes of the
energy-momentum tensor
\begin{gather}\label{T-xi-eta}
  T =\dd\xi\,\eta \,,
\end{gather}
and the fermionic screening is
\begin{equation*}
  S = \oint\eta(z) =\eta_0\,.
\end{equation*}
Via the mode expansions (in the integer-moded sector, which is the
analogue of the Neveu--Schwarz sector of ordinary free fermions)
\begin{gather}
  \xi(z)=\sum_{n\in\oZ}\xi_n\, z^{-n}\qquad{\rm and}\qquad
  \eta(z)=\sum_{n\in\oZ}\eta_n\, z^{-n-1} \,,
\end{gather}
the operator product \eqref{xi-eta} translates into
$\{\xi_m,\eta_n\}\,{=}\,\delta_{m+n,0}$.

We denote by $\Xi$ the $\xi$-$\eta$ module generated from the vacuum
state $\Vac$ that satisfies the annihilation conditions
\begin{equation}
  \eta_n\Vac=0 \quad{\rm for}\ \ n\,{\ge}\,0\,,\qquad
  \xi_n\Vac=0  \quad{\rm for}\ \ n\,{\ge}\,1\,.  
\end{equation}
The Virasoro algebra acts in $\Xi$ via~\eqref{T-xi-eta}.  With respect
to the $\mathfrak u(1)$ current $J\,{=}\,\eta\xi$, the $\eta$ field
(and hence all $\eta_n$) has charge $+1$ and $\xi$ (and hence all
$\xi_n$) has charge $-1$, while the energy-momentum tensor is neutral.
Because the Virasoro generators commute with $J_0$, we have the direct
sum decomposition
\begin{equation}\label{eq:2}
  \Xi=\bigoplus_{n\in\oZ}\Pi_n
\end{equation}
into indecomposable Virasoro modules $\Pi_n$ of fixed $\mathfrak u(1)$ 
charge $n$, the Feigin--Fuks (FF) modules. In view of the $\eta_0$ 
action, $\Xi$ is in fact a Felder complex of FF Virasoro modules.

\begin{figure}[tb]
  \begin{center}
    \includegraphics{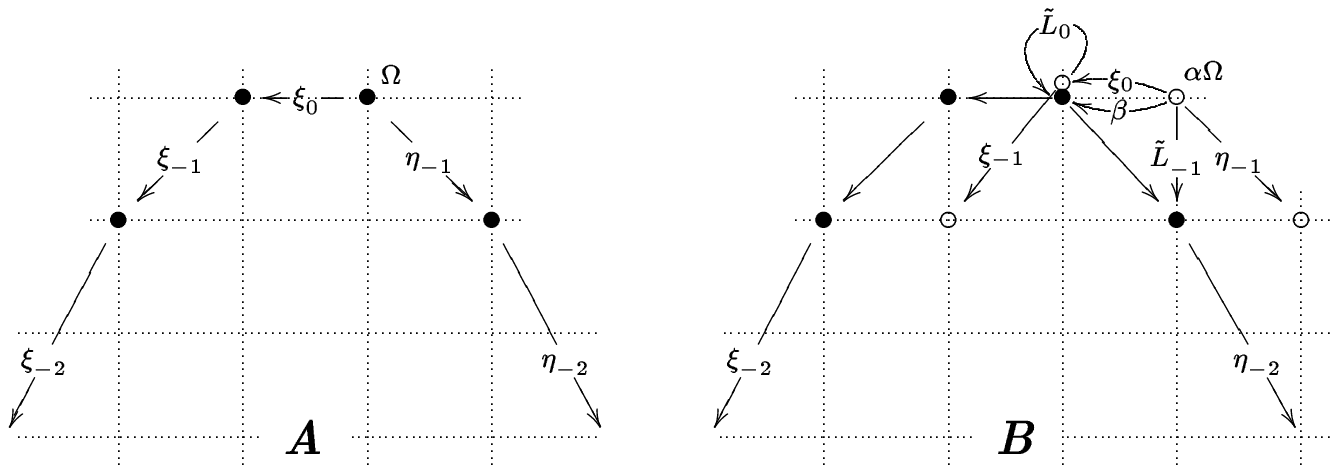}
  \end{center}
  \caption[The $\xi$-$\eta$ module]{\captionfont{A.\ The $\xi$-$\eta$
    module}.  Filled dots are the extremal states on the integer
    lattice corresponding to the (charge, level) coordinates on the
    plane.  Virasoro modules are arranged along vertical lines: each
    vertical line is a FF module.\\
    \captionfont{B.\ The ``logarithmically'' extended $\xi$-$\eta$
    module}. Open dots are descendant states of $\alpha\,\Vac$, the
    top-right open dot. Labels on some of the $\xi_n$ and $\eta_n$
    arrows are dropped for simplicity. Two instances of the Virasoro
    action on ``logarithmic'' states are shown.}
  \label{fig:xieta}
\end{figure}

The module $\Xi$ is shown in Fig.\;\ref{fig:xieta}\captionfont{A}.
States arranged along each vertical line furnish a FF module.  The
differential $\eta_0$ acts from a given FF module to the next module
on the right. The conformal weight of the highest-weight vector in
$\Pi_n$ is $\frac{n^2+n}{2}$. The structure of a FF module differs
from that of a Verma module by inverting a ``half'' of the submodule
embeddings. Each $\Pi_{n}$ with $n\,{<}\,0$ (to the left of the center
in Fig.~\ref{fig:xieta}\captionfont{A}) yields an irreducible Virasoro
module with conformal weight~$\frac{n^2+n}{2}$ as a \textit{quotient}
module (there is a nonvanishing singular vector at the highest
possible level built on the top-level state\,---\,the filled dot on
the corresponding vertical line). On the other hand, each $\Pi_{n}$
with $n\,{\geq}\,0$ (to the right of the center in
Fig.\;\ref{fig:xieta}\captionfont{A}) contains an irreducible Virasoro
representation as a \textit{sub}module (the ``first'' singular vector
vanishes, so that the corresponding top-level state is in a Virasoro
submodule of the module generated from a lower-lying state, a
``cosingular vector'').

In Fig.~\ref{fig:xieta}\captionfont{B}, we show the ``logarithmic''
extension of the $\xi$-$\eta$ module obtained via the prescription in
Sec.~\ref{sec:deform-general} with $E\,{=}\,\eta$. In accordance
with~\eqref{dA=E}, this amounts to introducing the state $\alpha\,\Vac$
(with $\beta\,\Vac\,{=}\,0$ and $\{\alpha,\beta\}\,{=}\,1$), 
which then corresponds to the field
\begin{align}\label{alpha-tilde}
  \tilde\alpha(z) &\equiv \dd^{-1}\eta(z)=\alpha+\eta_0\,\log z
  - \!{\sum_{n\in\,\oZ{\setminus}\{0\}}}\tfrac{1}{n}\,
  \eta_n\,z^{-n}\,.\\
  \intertext{The top-right open dot in
    Fig.~\ref{fig:xieta}\captionfont{B} is this new state
    $\alpha\,\Vac$. Next, the deformation of $\xi$ is given by}
\label{xi-tilde}
  \tilde\xi(z) &= (1{+}\beta\Delta_\eta)\xi(z)
  = \sum_{n\in\oZ} \xi_n\, z^{-n} + \beta\,\log z\,,
\end{align}
so that in particular $\tilde\xi(z)\tilde\alpha(w)
\,{=}\,{-}\log(z{-}w)$.  That the operator products are preserved by
$1{+}\beta\Delta_\eta$ (Theorem~\ref{thm:deform}) implies that the
deformed energy-momentum tensor
\begin{align*}
  \tilde T(z)={}&(1{+}\beta\Delta_\eta)T(z)
  =\dd\xi\,\eta(z) + \tfrac1z\,\beta\,\eta(z)\\
  \intertext{can be expressed in the form~\eqref{T-xi-eta} involving
    the deformed fields,}
  \tilde T(z)={}& \dd\tilde\xi(z)\,\eta(z)\,.
\end{align*}
It readily follows that acting by $\smash{\tilde{L}}_{-1}$ on
$\alpha\,\Vac$ yields the state $\eta_{-1}\Vac$ corresponding to the
operator~$\eta$.  Next, because of
\begin{equation}
  \tilde{L}_0\cdot\xi_0\alpha\,\Vac=\Vac\,,
\end{equation}
the state $\xi_0\alpha\,\Vac$ forms a Jordan cell with the vacuum.

\begin{Rem}
  We note that in our logarithmic deformation approach, $\beta$ can be
  viewed as a ``deformation parameter,'' whereas $\alpha$ is used in
  extending the space of states so as to maintain the operator--state
  correspondence. In~\eqref{alpha-tilde} and~\eqref{xi-tilde}, on the
  other hand, both $\alpha$ and $\beta$ enter as ``modes'' of the
  fields at $\log z$.  
\end{Rem}

The pair $(\tilde\xi,\tilde\alpha)$ of weight-$0$ fields constitute
the so-called symplectic fermions~\cite{kaus2,kaus3}; for the
symplectic fermions, the scalar product~$(~,~)_L^{}$ constructed in
Sec.~\ref{sec:Ward} becomes that in~\cite{gfn,kaus3}.  The
``logarithmic'' module $\modL$ generated by $\eta_n, \alpha, \xi_n,
\beta$ from $\Vac$ is an extension of~$\Xi$ by~$\Xi$,
\begin{equation}\label{J-sequence}
  0\longrightarrow\Xi\longrightarrow \modL
   \longrightarrow\Xi\longrightarrow0\,.
\end{equation}

To describe the corresponding extensions of \textit{Virasoro} modules,
we let $G_n$ denote the (usually reducible) module that is the
quotient of the Verma module with conformal weight $\frac{n^2-n}{2}$
over the singular vector at level $2n{+}1$. ($G_n$ is a submodule in
$\Pi_n$ generated from the extremal state with~$n>0$; the singular
vector at the level $2n{+}1$ vanishes in $\Pi_n$.)
\begin{Thm}
  As a Virasoro module, $\modL$ decomposes as
  \begin{equation}\label{J-decomp}
    \modL=\bigoplus_{n\geq0}\bigoplus_{m=0}^n \modL^n \,,
  \end{equation}
  where $\modL^n$ are nondiagonalizable modules obtained as the middle
  terms in the short exact sequences
  \begin{equation}\label{Jn-sequence}
    0\longrightarrow G_n\longrightarrow \modL^n\longrightarrow G_{n+1}
    \longrightarrow0\,,\qquad n\,{\geq}\,0\,.
  \end{equation}
\end{Thm}

To prove this claim, we first show that the right-hand side
of~\eqref{J-decomp} is embedded into $\modL$.  Consider the extremal
state $\ket n\,{\in}\,\modL$ with $n\,{>}\,0$.
The Virasoro module generated from $\ket n$ contains a singular vector
$\ket{n,n}$ at the level $n$.  The module $\modL^n$ is then generated
by the Virasoro algebra from the state $\ket{n}_L$ that corresponds to
the operator $\dd^{-1}\eta(z)\,\xi(z)$ evaluated on the state
$\ket{n,n}$.  For $n\,{=}\,0$, the module $\modL^0$ is generated from
the state
$\lim_{z\to0}\dd^{-1}\eta(z)\xi(z)\Vac\,{=}\,\alpha_0\xi_0\Vac$.  We
next use the fact that the operators
\begin{equation}\label{efh}
  e =\alpha_0\eta_0+\sum_{n\neq0} \tfrac{1}{n}\,\eta_{-n}\eta_n\,,
  \qquad
  f=\xi_0\beta_0+\sum_{n\in\oZ} n\,\xi_{-n}\xi_n\,,\qquad
  h = J_0+\alpha_0\beta_0
\end{equation}
span the Lie algebra $\sL(2)$. This algebra acts in $\modL$ and
commutes with the Virasoro algebra represented via the modes
$\tilde{L}_n$ of~$\tilde T(z)$. We can therefore employ it to carry
over the modules generated from the above states, ``at the edge'' of
$\Xi$, into the ``interior'' of the module.  Specifically, because
$e\ket{n}_L\,{=}\,0$ and~$f^{n+1}\ket{n}_L\,{=}\,0$, taking the
$\sL(2)$ orbit through~$\ket{n}_L$ gives the embedding of
$\bigoplus_{m=0}^n\modL^n$ into $\modL$.
   
The theorem now follows by comparing the characters of the left- and
the right-hand sides in~\eqref{J-decomp} to show that the $\SL2$
orbits exhaust $\modL$.  The character of the left-hand side is
\begin{equation}\label{eq:6}
  (1+z)\prod_{i\geq0}(1+z^{-1}q^i)\prod_{i\geq1}(1+zq^i) \,,
\end{equation}
while the character of the right-hand side is
\begin{equation}\label{eq:7}
  \frac{1}{\prod_{i\geq1}(1-q^i)}\Bigl(
  \sum_{n\geq0}\sum_{m=0}^n z^{n-2m}\bigl(
  q^{\frac{n^2-n}{2}}(1-q^{2n+1})
  +q^{\frac{n^2+n}{2}}(1-q^{2n+3})\bigr) \Bigr) \,.
\end{equation}
These coincide because of the Jacobi triple product
identity.\hfill\rule{1.5ex}{1.5ex}

\medskip

Because $\Xi$ is a direct sum of the $\Pi_n$ modules, the
sequence~\eqref{J-sequence} decomposes into a collection of short
exact sequences
\begin{equation}\label{FJF-sequence}
  0\longrightarrow \Pi_{n}\longrightarrow \modL_n\longrightarrow
  \Pi_{n-1}
  \longrightarrow0\,,\qquad n\in\oZ\,,
\end{equation}
where $\modL_n$ is the component with a definite charge in $\modL$,
$\modL\,{=}\,\bigoplus_{n\in\oZ}\modL_n$. We then have
\begin{equation}
  \modL_n=\bigoplus_{m\geq n}\modL^n
\end{equation}
(that the sum is direct again follows by making use of the $\sL(2)$
algebra generated by operators~\eqref{efh}).

\subsection{The $c\,{=}\,0$ theory and the $\N2$ superconformal
  algebra}
\label{sec:N2} 
The (trivial) $c\!=\!0$ Virasoro minimal model is naturally embedded
into the $\N2$ algebra in such a manner that the fermionic screening
becomes the zero mode of one of the two fermionic
currents.\footnote{~Even in the trivial model, it is instructive to
  see how Jordan cells open in a free-field realization of the
  reducible Virasoro modules entering the Felder complex.  In the
  logarithmic context, a reasonable object to study is thus the
  quotient of the vacuum Verma module by the submodule generated from
  its first singular vector $L_{-1}\Vac$~\cite{mila3}.  We here deal
  with a free-field counterpart of this quotient module.}  For this,
we take the $\N2$ superconformal algebra in the topologically
twisted~\cite{egya2} basis where the two fermionic fields have the
weights~$1$ and~$2$,
\begin{align*}
  T(z)\,T(w)&=\frac{2T(w)}{(z-w)^2}+\frac{\dd T}{z-w}\,,&
  T(z)\,H(w)&=\frac{c/3}{(z-w)^3}+\frac{H(w)}{(z-w)^2}+\frac{\dd H}{z-w}\,,\\
  H(z)\,H(w)&=\frac{c/3}{(z-w)^2}\,, &
  Q(z)\,G(w)&=\frac{c/3}{(z-w)^3}-\frac{H(w)}{(z-w)^2}+\frac{T}{z-w} \,,\\
  T(z)\,G(w)&=\frac{2G(w)}{(z-w)^2}+\frac{\dd G}{z-w}\,,&
  T(z)\,Q(w)&=\frac{Q(w)}{(z-w)^2}+\frac{\dd Q}{z-w}\,,\\  
  H(z)\,G(w)&=\frac{G}{z-w}\,,&
  H(z)\,Q(w)&=-\frac{Q}{z-w}\,.
\end{align*}
Here, $T$ is the energy-momentum tensor, $H$ is the $\mathfrak u(1)$
current, and~$G$ and~$Q$ are the fermionic fields of the respective
conformal weights~$2$ and~$1$.  

With $E\,{=}\,Q$, we calculate $\tilde A\,{=}\,(1{+}\beta\Delta_Q)A$
using deformation formula~\eqref{Delta-E}, with the result
\begin{align}
  \tilde H &=  H + \beta\log z\; Q\,,\\
  \tilde T &=  T + \tfrac{\beta}{z}\, Q\,,\\
  \tilde G &= G + \beta\log z\,T
  - \tfrac{\beta}{z}H - \beta\tfrac{c}{6 z^2}\,.
\end{align}
Although we need the $\N2$ algebra representations with the $\N2$
central charge $c\,{=}\,1$, the above deformation formulas are valid
for an arbitrary value of~$c$.

For $c\,{=}\,1$, there are three irreducible unitary $\N2$ modules.
Each of them decomposes into a direct sum $\bigoplus_{n\in\oZ}\Pi_n$
of FF modules over the Virasoro algebra with zero central charge.  The
modules $\Pi_n$ are distinguished by the action of the zero mode $H_0$
of the $\mathfrak u(1)$ current (with $H_0$ acting as multiplication
by $n$ on $\Pi_n$ in the case of the vacuum $\N2$ module).  Moreover,
the Virasoro modules sitting inside a unitary $\N2$ module for $k=1$
arrange into the Felder complex for the \textit{trivial}
module.\footnote{~In general, the irreducible Virasoro module (in the
  $(2,3)$ case, the trivial one) is selected by \textit{two} Felder
  complexes that correspond to two different screenings.  In the
  $(2,3)$ case, these are $Q_0\,{=}\oint\! \rme^{\sqrt{3}\varphi}$
  and $\oint \!\rme^{-\frac{2}{\sqrt{3}}\varphi}$.}

\begin{figure}[tb]
  \begin{center}
    \includegraphics[trim= 0 500 0 0, clip]{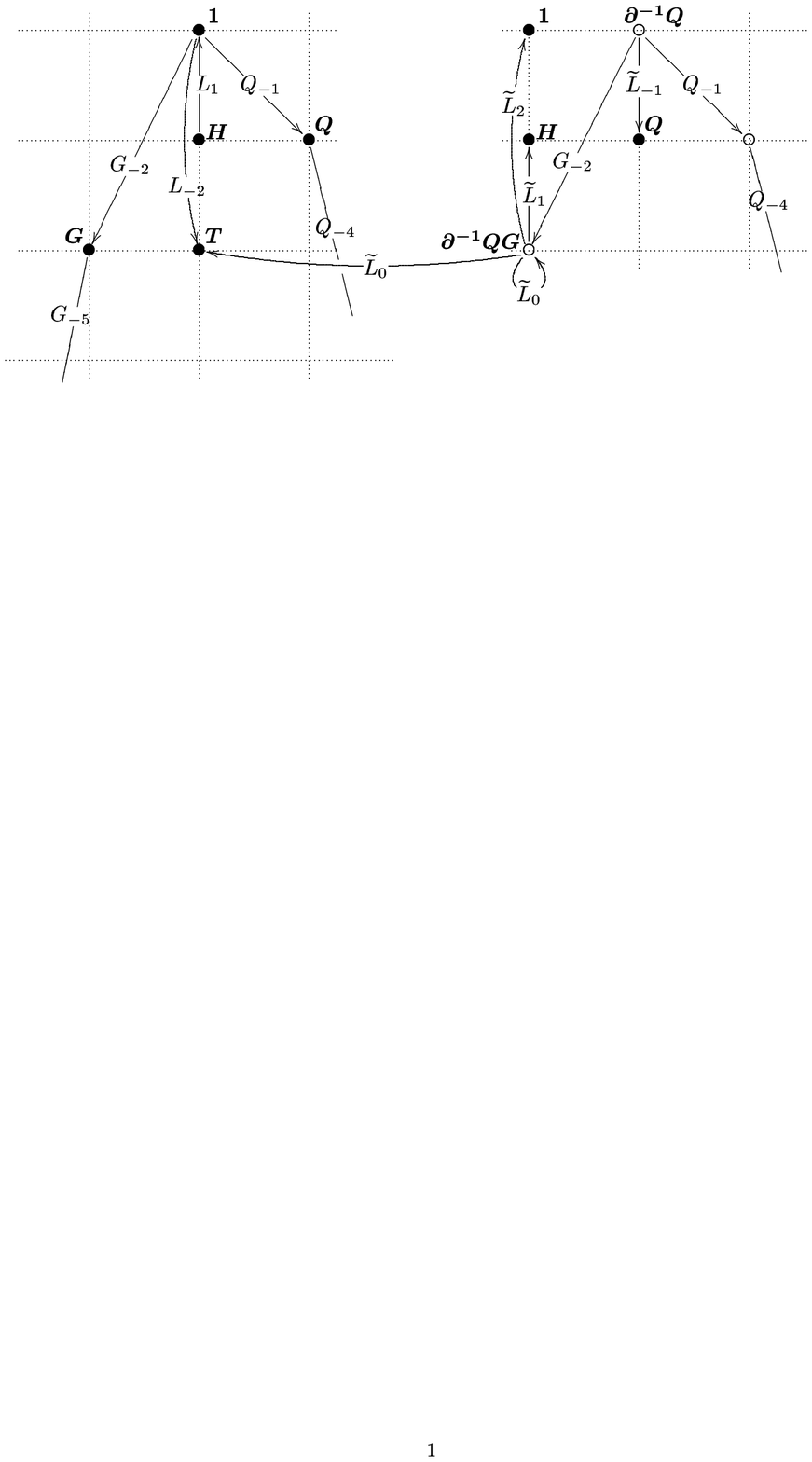}
  \end{center}
  \caption[A logarithmically extended Virasoro module inside
  unitary $\N2$ module]{\captionfont{A logarithmically extended
      $c\,{=}\,0$ Virasoro module inside the $\dd^{-1}\!Q$-extended
      unitary $\N2$ module}.  Filled dots are states in the vacuum
    $k\,{=}\,1$ unitary module of the $\N2$ algebra.  The integer
    lattice corresponds to the (charge, level) coordinates.  Shown in
    boldface are the \textit{operators} corresponding to the states.
    In the right part, open dots depict the ``logarithmic'' states,
    the $\N2$ descendants of $\dd^{-1}\!Q$.  The right part must be
    superimposed on the left one such that double occurrences of
    identical states ($\Oneshow$, $\Hshow$, and $\Qshow$) coincide.
    The states corresponding to the operators $T$ and $\dd^{-1}Q G$
    are then in the same grade and produce a Jordan cell for the
    Virasoro generator $\tilde{L}_0$.}
  \label{fig:N2}
\end{figure}

In Fig.~\ref{fig:N2}, we display the structure of the vacuum $\N2$
module (which contains the vacuum Virasoro module) and its
``logarithmic'' extension.
The left part of the picture shows states in the $k\,{=}\,1$ vacuum
unitary module, on which the operator constraints
\begin{gather}\label{Q-G-identities}
  \dd Q\,Q=0\qquad \dd G\,G=0
\end{gather}
are valid.\footnote{~In fact, whenever $c\,{=}\,3k/(k+2)$ with
  $k\,{\in}\,\oN$, the unitary representations of the $\N2$ algebra
  are characterized by the operator constraints
  $\dd^{k}Q\,\dd^{k-1}Q\cdots\dd Q\,Q=0$ and
  $\dd^{k}G\,\dd^{k-1}G\cdots\dd G\,G=0$~\cite{semi-inf}.}  This in
particular determines the ``shape'' of the module (e.g., the vacuum is
annihilated by $Q_{\geq0}$ and $G_{\geq-1}$, etc.).  With boldface
symbols, we indicate the operators that correspond to the states (for
example, the state corresponding to $H$ is a Virasoro
\textit{cosingular} vector with respect to the vacuum state).

The right part of the picture shows the ``logarithmic'' states
generated from $\alpha\,\Vac$, where $\{\alpha,\beta\}\,{=}\,1$ (with
$\beta\,\Vac\,{=}\,0$), and $\alpha\,\Vac$ therefore corresponds to 
the operator $\dd^{-1}Q$ in accordance with~\eqref{dA=E}. 
The right part of the picture must be superimposed onto the left one
such that the two copies of each state displayed in both parts
($\Oneshow$, $\Hshow$, and~$\Qshow$) coincide.  Again, the statement
that $1{+}\beta\Delta_Q$ is an isomorphism of operator products
implies that we can ``hide'' $\beta$ into the deformed operators and
$\alpha$ into $\dd^{-1}Q$ and consistently work with the deformed
(tilded) operators. The action of the modes $\tilde{L}_n$ on the
states in the $\N2$ module generated from $\alpha\,\Vac$ then relates
these states to states in the original modules~$\Pi_n$.  In the
logarithmically extended vacuum Virasoro module, in particular, the
first Jordan cell opens at level~$2$; it is spanned by the states
$\tilde{L}_{-2}\Vac=L_{-2}\Vac$ and $G_{-2}\alpha\Vac$ that correspond
to the respective operators~$\tilde{T}$ and~$\dd^{-1}Q\,\tilde G$,
\begin{align}
  [\tilde L_0, \dd^{-1}Q\,\tilde G(z)] &=
  2 \dd^{-1}Q\,\tilde G(z) + \tilde T(z)\,.\\
 \intertext{We also have the positive-mode Virasoro action on the
 ``Jordan'' state}
  [\tilde L_1, \dd^{-1}Q\,\tilde G(z)] &= -\tilde H(z)\,,\\
  [\tilde L_2, \dd^{-1}Q\,\tilde G(z)] &= \tfrac{c}{3}\,.
\end{align}

We note that a realization of $Q$ and $G$ satisfying
relations~\eqref{Q-G-identities} is given by
$Q\,{=}\,\rme^{\sqrt{3}\varphi}$ and
$G\,{=}\,\rme^{-\sqrt{3}\varphi}$, where $\varphi$ is a scalar field
with the OPE
\begin{equation*}
  \dd\varphi(z)\,\dd\varphi(w)=\frac{1}{(z-w)^2}\,.
\end{equation*}
The energy-momentum tensor then becomes $T\,{=}\,\half\,\dd\varphi
\dd\varphi +\frac{1}{2\sqrt{3}}\,\dd^2\varphi$.

\medskip

Regarding the $\N2$ superconformal algebra, we also note that its
deformation via $E\,{=}\,H$ produces (as a particular case
of~\eqref{spflow-gen1}--\eqref{spflow-gen3}) the familiar spectral
flow transform~\cite{schwimseib}, which is an automorphism for
$\beta\in\oZ$, while the deformation via the weight-$2$ fermionic
field does not seem to admit an interesting interpretation.
  

\subsection{$(2,p)$ minimal models}
\label{sec:W} 
The cases considered above are generalized as follows.  Adding the
screening current to the Virasoro algebra with central charge
\begin{gather}\label{2pcc}
  c_p=1-3\,\tfrac{(p-2)^2}{p}
\end{gather}
of the $(2,p)\,{=}\,(2,2r{+}1)$ minimal model gives the algebra
generated by fermionic currents $E(z)$ and $F(z)$ satisfying the
additional constraints (cf.~\eqref{Q-G-identities})
\begin{align}
  \dd E(z)\,E(z)&=0\,,\qquad\dd^3\! E(z)\,E(z)=0\,,\quad
  \dots\,,\quad\dd^{p-2}\!E(z)\,E(z)=0\,,\\
  \dd F(z)\,F(z)&=0\,,\qquad\dd^3\! F(z)\,F(z)=0\,,\quad
  \dots\,,\quad\dd^{p-2}\!F(z)\,F(z)=0
\end{align}
and the OPE
\begin{equation}\label{eq:EF-to-H}
  E(z)F(w)=\frac{1/p}{(z-w)^{p}} +\frac{H^0}{(z-w)^{p-1}}
  +\frac{H^1}{(z-w)^{p-2}}+\,\dots\, +\frac{H^{p-2}}{z-w}\,.
\end{equation}
The fields involved here generate a W-algebra $\cA_{(p)}$; the lowest
generators are the $\mathfrak u(1)$ current $J\,{=}\,H^0$ and the
energy-momentum tensor $T\,{=}\,H^1\,{+}\,\frac{p-3}{2}\,\dd H^0$ with
central charge~\eqref{2pcc}. The currents $F$ and $E$ are then
primary fields with the respective conformal weights~$p{-}1$ and~$1$.
Thus the mode decompositions are given by
\begin{align}
  E(z)=\sum_{n\in\oZ}E_n\,z^{-n-1}\,,\quad\ 
  F(z)=\sum_{n\in\oZ}F_n\,z^{-n-p+1}\,,\quad\ 
  H^m(z)=\sum_{n\in\oZ}H^m_n\,z^{-n-m-1}\,.
\end{align}
In particular, the screening is given by~$E_0$.

We now let $\Xi_n$, with $n\,{=}\,0,1,...\,,p{-}1$, denote the
$\cA_{(p)}$-module generated from the state $\ket{n}$ satisfying
\begin{equation}
  \begin{split}
    E_i\ket{n}={}&0\qquad {\rm for}\quad i\,{\geq}\, n\,,\\
    F_j\ket{n}={}&0\qquad {\rm for}\quad j\,{\geq}\,{-}p{-}n{+}2\,.
  \end{split}
\end{equation}
For all $m\,{\in}\,\oZ$, the states $\ket{n+m p}$ then belong 
to~$\Xi_n$.  We call~$\Xi_0$ the vacuum module.

As a \textit{Virasoro} module, each $\Xi_n$ decomposes into the direct
sum
\begin{equation}\label{eq:xi-decomposition}
  \Xi_n=\bigoplus_{m\in\oZ}\Pi_{\frac{n}{p}+m}
\end{equation}
of FF modules, where the subscript indicates that $H^0_0$ acts as 
multiplication by $h$ on $\Pi_h$. Each~$\Pi_{\frac{n}{p}+m}$ is
generated by~$H^0_\ell$ with $\ell\,{<}\,0$ from the 
state~$\ket{n{+}pm}$,
and $\Xi_n$ is a Virasoro Felder complex with the differential~$E_0$.

In the vacuum Virasoro module $\Pi_0$, the current $H^0$ generates a
cosingular vector and $H^{p-2}$ a singular vector.  The latter
determines the position of the first Jordan cell in the corresponding
logarithmic vacuum module obtained as the extension
\begin{equation}
  0\to\Pi_0\to\modL(0)_0\to\Pi_{1}\to0\,,
\end{equation}
as we now explain.  Following the general prescription, we consider
the currents $\tilde A\,{:=}\,A\,{+}\,\beta\Delta_E A$, where
relations~\eqref{alpha-kill} and~\eqref{alpha-beta-1} hold for $\alpha$
and $\beta$, which implies that the state $\alpha\ket{0}$ corresponds
to ${-}\dd^{-1}\!E$. The states corresponding to $\dd^{-1}\!E\tilde F(z)$
and to $\tilde H^{p-2}$ then span a Jordan cell, because
\begin{alignat}{2}
  [\tilde L_0,\dd^{-1}\!E\tilde F(z)]&=(p\,{-}\,1)\dd^{-1}\!E
  \tilde F(z)+ \tilde H^{p-2}\,.\\
 \intertext{We also have the positive-mode Virasoro action on the
 ``Jordan'' state that corresponds to the operator $\dd^{-1}\!E\tilde
    F(z)$,} [\tilde L_n,\dd^{-1}\!E\tilde F(z)]&=\tilde
  H^{p-n-2}(z)&&\kern-40pt
  {\rm for}\ \ 1\,{\leq}\,n\,{\leq}\,p\,{-}\,2  
  \,,\\
  [\tilde L_{p-1},\dd^{-1}\!E\tilde F(z)]&=\tfrac{1}{p}\,,
\end{alignat}
and $[\tilde L_n,\dd^{-1}\!E\tilde F(z)]\,{=}\,0$ for $n\,{\geq}\,p$.

This result therefore generalizes the picture in Fig.~\ref{fig:N2}:
the module becomes ``narrower'' than in Fig.~\ref{fig:N2} because
instead of $G_{-2}$, the first mode of $F$ that does not annihilate
the vacuum is $F_{-p+1}$.  We have $p$ modules $\modL(n)$ that
correspond to the state space of the VOA~$\tilde\voaV'$ introduced
after the proof of Lemma~\ref{thm:general}.  Each of them can be
obtained as the extension
\begin{equation}
  0\to\Xi_n\to\modL(n)\to\Xi_n\to0
\end{equation}
of Virasoro modules.  At the same time, each module~$\Xi_n$ decomposes
into the direct sum~\eqref{eq:xi-decomposition} of modules, and
the previous exact sequence is in fact the direct sum of the sequences
\begin{equation}
  0\to\Pi_{\frac{n}{p}+m}\to\modL(n)_m\to\Pi_{\frac{n}{p}+m+1}\to0\,,
\end{equation}
where $\Xi_n\,{=}\,\bigoplus_m\Pi_{\frac{n}{p}+m}$ and
$\modL(n)\,{=}\,\bigoplus_m\modL(n)_m$.

We also note that in the one-scalar realization, the energy-momentum
tensor is given by
\begin{equation}\label{eq:the-emt}
  T =\half\,\dd\varphi\,\dd\varphi +\frac{p-2}{2\sqrt{p}}\,\dd^2\varphi
  \,,
\end{equation}
and the fields $E$ and $F$ are
\begin{equation}\label{eq:screening-and-conjugated}
  E=\tfrac{1}{\sqrt{p}}\,\rme^{\sqrt{p}\varphi}\,,\qquad
  F=\tfrac{1}{\sqrt{p}}\,\rme^{-\sqrt{p}\varphi}\,,
\end{equation}
with the screening operator for $T$ given by
$\oint\!\rme^{\sqrt{p}\varphi}$.
The states $\ket{n}$ then correspond to fields as
\begin{gather}\label{eq:extremal-states}
  \ket{n}=\left.\rme^{-\frac{n}{\sqrt{p}}\varphi(z)}\Vac
  \right|_{z\to0}\,.
\end{gather}

\section{A further deformation example: logarithmic $\protect\hSL2$}
\label{sec:log-sl2}
We continue applying the above strategy to deform a vertex operator
algebra using higher-order poles that its fields develop with the
integrand of a fermionic screening operator.  For the $\hSL2$ algebra
with a fermionic screening, this results in constructing
indecomposable $\hSL2$ modules by ``gluing together'' two modules
(i.e., extending one module with the other) that are adjacent in the
resolution with the differential given by the fermionic screening.
This resolution involves \textit{twisted} (spectral-flow transformed)
$\hSL2$-modules.

The $\hSL2$ algebra is generated by three currents with the OPEs
\begin{align*}
    J^0(z)J^\pm(w)&={}\pm\frac{J^\pm}{z-w}\,,
    \qquad J^0(z)J^0(w)={}\frac{k/2}{(z-w)^2}\,,\\
    J^+(z)J^-(w)&={}\frac{k}{(z-w)^2}+\frac{2J^0}{z-w}\,.
\end{align*}
Spectral flow with the transformation parameter $\theta$ acts on the
modes introduced via $J^{\pm,0}(z)=\sum_{n\in\oZ}J^{\pm,0}_n z^{-n-1}$
as
\begin{equation*}
  J^{\pm}_n\mapsto J^{\pm}_{n\pm\theta}\,,\qquad
  J^0_n\mapsto J^0_n + \tfrac{k\theta}{2}\,\delta_{n,0}\,.
\end{equation*}
For $\theta\,{\in}\,\oZ$, this mapping is an automorphism of the 
$\hSL2$ algebra.

\subsection{A fermionic screening for $\protect\hSL2$}
\label{sec:ferm-scr-l2}
We describe the level-$k$ $\hSL2$ current algebra as constructed from
the Virasoro algebra with central charge
\begin{equation}\label{eq:d(k)}
  d = 13 -\frac{6}{k+2} - 6(k+2)
\end{equation}
and two free boson fields $\varphi$ and $\phi$ with the operator
products
\begin{equation}
  \dd\varphi(z)\,\dd\varphi(w)=\frac{1}{(z-w)^2}\,,\qquad
  \dd\phi(z)\,\dd\phi(w)=\frac{-1}{(z-w)^2}\,.
\end{equation}
The $\hSL2$ currents are given by~\cite{S-sing}
\begin{align}
  J^+ ={}& \rme^{\sqrt{\frac{2}{k}}(\varphi - \phi)}\,,\label{Jplus}\\
  J^0 ={}& \sqrt{\tfrac{k}{2}}\,\dd\varphi\,,\label{Jnaught}\\
  J^- ={}& (k{+}2)\, T\,\rme^{-\sqrt{\frac{2}{k}}(\varphi - \phi)} -
  \tfrac{k}{2}\, \dd\phi\, \dd\phi\,\,
  \rme^{-\sqrt{\frac{2}{k}}(\varphi - \phi)}
  - \sqrt{\tfrac{k}{2}}\,(k{+}1)\, \dd^2\phi\,
  \rme^{-\sqrt{\frac{2}{k}}(\varphi - \phi)}\,,
  \label{Jminus}
\end{align}
where $T\,{=}\sum_{n\in\oZ}L_n z^{-n-2}$ is the energy-momentum tensor
corresponding to the Virasoro generators with central
charge~\eqref{eq:d(k)}. The Sugawara energy-momentum tensor, of
central charge $c_{\rm Sug}\,{=}\,\frac{3k}{k+2}$, then evaluates as
\begin{gather}\label{eq:Sug}
  T_{\mathrm{Sug}} = T + \thalf\,\dd\varphi\,\dd\varphi -
  \thalf\,\dd\phi\,\dd\phi - \sqrt{\tfrac{k}{2}}\,\dd^2\phi\,.
\end{gather}

A (fermionic) screening operator for the $\hSL2$ realization in
Eqs.~\eqref{Jplus}--\eqref{Jminus} is given by
\begin{equation}
  S = \oint \rme^{\sqrt{\frac{k}{2}}\phi}\,V_{12} \equiv \oint E(z)\,,
\end{equation}
where $V_{r,s}$ are Virasoro primary fields of the conformal weights
\begin{equation}  
  h_{r,s}=\frac{r^2-1}{4(k+2)}+\frac{s^2-1}{4}(k+2)+\frac{1-rs}{2}\,.
\end{equation}
In particular, $V_{12}$ and $V_{21}$ have the respective weights
\begin{equation}
  h_{12} = \tfrac{1}{4}\,(3(k+2) - 2)\qquad{\rm and}\qquad
  h_{21} = \tfrac{1}{4}\,(\tfrac{3}{k+2} - 2)\,.
\end{equation}
Because of singular vector decoupling, they satisfy the differential
equations
\begin{equation}\label{V12-diff}
  \dd^2 V_{12}(z) - (k+2)\,{\boldsymbol{:}}T(z)\,V_{12}(z){\boldsymbol{:}}
  = 0\,,
  \qquad
  \dd^2 V_{21}(z) -
  \tfrac{1}{k+2}\,{\boldsymbol{:}}T(z)\,V_{21}(z){\boldsymbol{:}}
  = 0\,.
\end{equation}
That $S$ is a screening follows from the differential equation
satisfied by $V_{12}$\,: the first-order pole of the OPE \ 
$E(z)\,J^-(w)$ is evaluated to be proportional to $\dd^2 V_{12}(z) -
(k{+}2)\,T(z)\,V_{12}(z)$.

$\hSL2$ primary fields are constructed as
\begin{align}
  \Phi_{r,s} &= \rme^{j(r,s)\sqrt{\frac{2}{k}}(\varphi-\phi)}\,V_{r,s}\\
  \intertext{with}
  j(r,s) &= \tfrac{r-1}{2} - \tfrac{s-1}{2}\,(k+2)\,.
\end{align}
For positive integers $r$ and $s$, singular vectors in the $\hSL2$
module generated from $\Phi_{r,s}$ are proportional to Virasoro
singular vectors, and once $V_{r,s}$ is the vertex operator
corresponding to an \textit{irreducible} Virasoro module, the
corresponding $\hSL2$ module is also irreducible.  We let
$\ket{\Phi_{r,s}}$ denote the highest-weight state corresponding to
the operator $\Phi_{r,s}$ (in particular,
$\ket{\Phi_{1,1}}\,{=}\,\Vac$ is the vacuum state) and let
$\modM_{r,s}$ be the module generated from $\ket{\Phi_{r,s}}$ by the
modes of the currents in Eqs.~\eqref{Jplus}--\eqref{Jminus}.

\begin{Rem}
  If the energy-momentum tensor $T$ is bosonized through a scalar
  current $\dd u$, the three currents $\dd\varphi$, $\dd\phi$, and
  $\dd u$ can be mixed and ``debosonized'' into $\beta$-$\gamma$ ghost
  fields (a first-order \textit{bosonic} system) and a current such
  that the $\hSL2$ currents in Eqs.~\eqref{Jplus}--\eqref{Jminus}
  become those of the Wakimoto realization~\cite{waki}.  The fermionic
  screening then becomes $S\,{=}\,\eta_0$, the zero mode of the
  $\xi$-$\eta$ system entering the bosonization of the $\beta$ and
  $\gamma$ fields~\cite{frms}.  The logarithmic representation of
  $\hSL2$ constructed with the use of the fermionic screening is
  therefore related to the symplectic fermions, see
  Sec.~\ref{sec:c=-2}.
  
  We also note that the realization in
  Eqs.~\eqref{Jplus}--\eqref{Jminus} is useful because under the
  mapping of the Wakimoto bosonization back to the $\dd u$,
  $\dd\varphi$, and $\dd\phi$ fields, the other two Wakimoto
  bosonization screenings are expressed solely through $u$, and
  therefore single out precisely the Virasoro algebra out of the
  free-scalar sector; these screenings therefore do not play a role as
  long as the $\hSL2$ currents are expressed through the
  energy-momentum tensor~$T$, as is the case with
  Eqs.~\eqref{Jplus}--\eqref{Jminus}.
\end{Rem}

\subsection{The deformation and indecomposable representations}
We now construct an indecomposable $\hSL2$ module in which
$\modM_{r,s}$ is a submodule.  The nonzero poles in the OPEs of $E$
with the algebra currents are given by
\begin{align}
  \opepole[2][E,J^-]&=
  \bigl(k\,\dd V_{12} +(k{+}2)\sqrt{\tfrac{k}{2}}\,\dd\phi\,
  V_{12}\bigr)\,
  \rme^{\sqrt{\frac{k}{2}}\phi + \frac{2}{k}(\phi-\varphi)}\,,\\
  \opepole[3][E,J^-]&=2(k{+}1)\,
  \rme^{\sqrt{\frac{k}{2}}\phi + \frac{2}{k}(\phi-\varphi)}
  \,V_{12}\,.
\end{align}
Thus the $J^-$ current is deformed as
\begin{equation}
  \tilde{J}^-(z) = (1{+}\beta \Delta_E)J^-(z)
  =J^-(z) + \beta\,\frac{\opepole[2][E,J^-]\!(z)}{z}
  - \beta\,\frac{\opepole[3][E,J^-]\!(z)}{2z^2}\,,
\end{equation}
while $J^+$ and $J^0$ remain unchanged.  The Sugawara energy-momentum
tensor is deformed as in Sec.~\ref{sec:deform-general},
\begin{equation}
  \tilde{T}_{\mathrm{Sug}}(z)
  = T_{\mathrm{Sug}}(z) +\tfrac{\beta}{z}\,E(z).
\end{equation}
In particular, $\tilde{L}_0\,{=}\,L_0{+}\beta E_0$ acts on states as
$\tilde{L}_0\ket{x}=L_0\ket{x} + \beta S\ket{x}$.

We next define the vertex operator 
\begin{equation}
  E^*_{r,s}=
  \rme^{-\sqrt{\frac{k}{2}}\phi +
    j(r,s)\sqrt{\frac{2}{k}}(\varphi-\phi)}
  \,V_{r,s-1}
\end{equation}
and let $\ket{S^*_{r,s}}$ denote the corresponding state.  It is a
``partner'' to the highest-weight state because
\begin{equation}
  \beta S\cdot\alpha\ket{S^*_{r,s}}=\ket{\Phi_{r,s}},
\end{equation}
and therefore, $\alpha\ket{S^*_{r,s}}$ and $\ket{\Phi_{r,s}}$
constitute a Jordan cell for the $\tilde{L}_0$ generator.  In
particular, $\beta S\,{\cdot}\,\alpha\ket{S^*_{1,1}}\,{=}\,\Vac$.
This partner to the vacuum state satisfies the highest-weight
conditions
\begin{equation}
  J^-_0\ket{S^*_{1,1}}=0\qquad{\rm and}\qquad
  J^+_{1}\ket{S^*_{1,1}}=0\,,
\end{equation}
and is therefore the $\theta{=}1$ spectral flow transform of a state
that satisfies the standard highest-weight conditions
($J^-_1\,{\approx}\,0$, $J^+_{0}\,{\approx}\,0$). For $rs\,{>}\,1$,
 both $J^-_0$ and $J^+_0$ are nonvanishing on $\ket{S^*_{r,s}}$.

\begin{figure}[tb]
  \begin{center}
    \includegraphics[trim= 0 550 0 -10, clip]{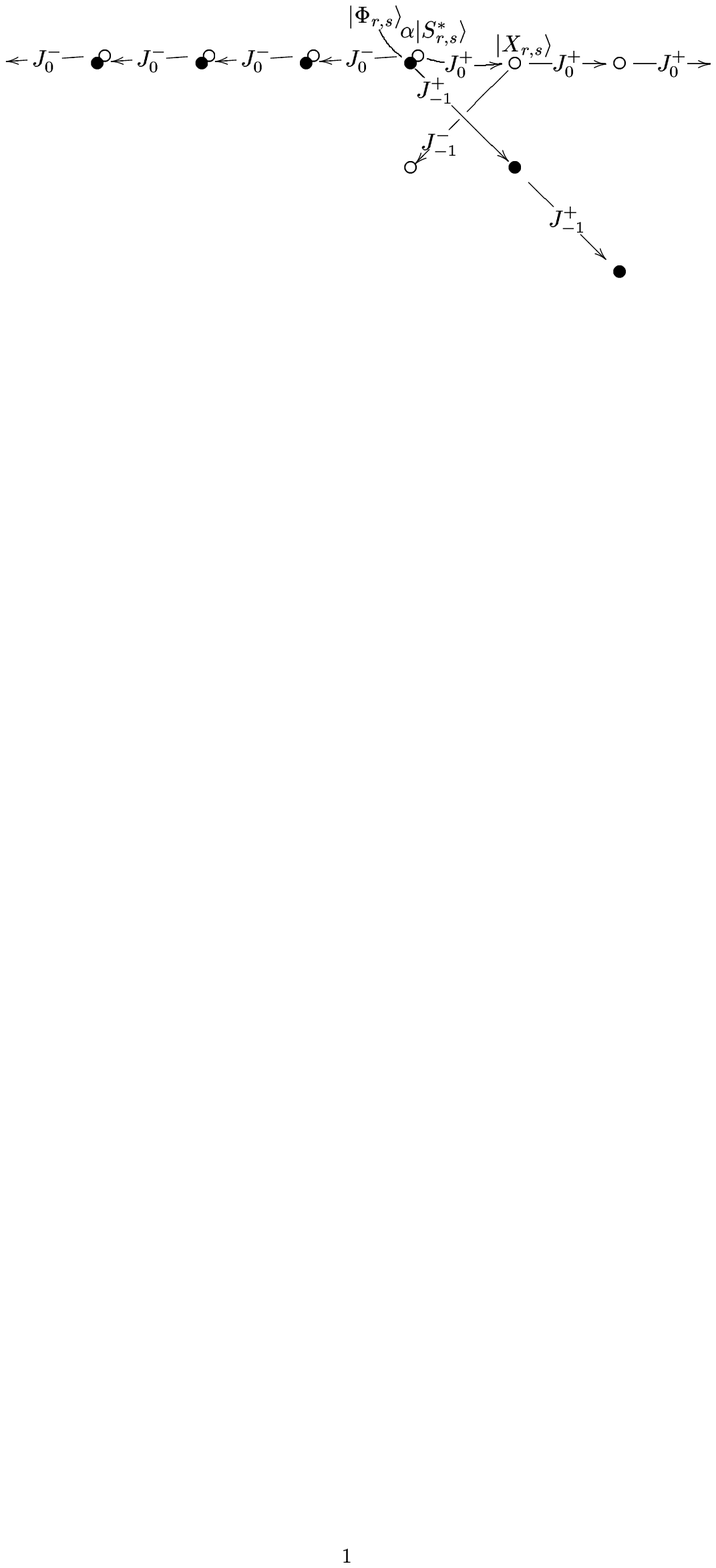}
  \end{center}
  \caption[An indecomposable extension of $\hSL2$
  modules]{\captionfont{An indecomposable extension of $\hSL2$
      modules.}  Filled dots are the extremal states in~$\modM_{r,s}$,
    with the top-right one being the highest-weight vector.  Open dots
    are states generated from $\alpha\ket{S^*_{r,s}}$. The partner
    $\alpha\ket{S^*_{r,s}}$ to the highest-weight vector
    $\ket{\Phi_{r,s}}$ does not satisfy the standard highest-weight
    conditions; however, these are recovered in the quotient module
    with respect to the submodule generated from $\ket{X_{r,s}}$.  It
    is understood that the states are at the sites of an integer
    lattice, with the charge axis directed to the right and the level
    (conformal weight) axis downward.}
  \label{fig:sl2}
\end{figure}

The action of $(J^+_0)^n$, $n\,{\geq}\,1$, on $\alpha\ket{S^*_{r,s}}$
generates states in the grades where the module $\modM_{r,s}$ contains
no states (the horizontal line of open dots in Fig.~\ref{fig:sl2} to
the right of the highest-weight vector).\footnote{~Pictures in this
  section have a somewhat different meaning than in the previous
  section.  There, we showed a module of the larger algebra, with each
  Virasoro module occupying merely a vertical line.  Here, we draw the
  analogue of that vertical line on the plane, whereas showing the
  module of the larger algebra would require a third dimension (the
  charge with respect to the $\phi$ boson).}
However, the very first of these states,
$\ket{X_{r,s}}\,{=}\,J^+_0\alpha\ket{S^*_{r,s}}$, is a
(spectral-flow-transformed) singular vector in the module
$\Sigma_{r,s}$ generated from $\alpha\ket{S^*_{r,s}}$: it satisfies
the twisted highest-weight conditions
\begin{equation}
  J^-_0\,\ket{X_{r,s}}=0\,,\qquad J^+_{1}\,\ket{X_{r,s}}=0\,.
\end{equation}
These ensure that a submodule $\modX_{r,s}$ is generated from
$\ket{X_{r,s}}$.  Taking the quotient $\Sigma_{r,s}/\modX_{r,s}$
eliminates all states in the $(\mathrm{level},\mathrm{charge})$-grades
$(\ell,h)$ with $h\,{>}\,\ell$ (i.e., in the grades ``outside'' the
$\modM_{r,s}$ module).  The module $\modL_{r,s}$ constructed this way
is a natural ``logarithmic'' indecomposable extension of~$\modM_{r,s}$,
\begin{equation*}
    0\to\modM_{r,s}\to\modL_{r,s}\to\Sigma_{r,s}/\modX_{r,s} \to0\,.
\end{equation*}

\begin{figure}[tb]
  \begin{center}
    \includegraphics[trim= 0 550 0 -10, clip]{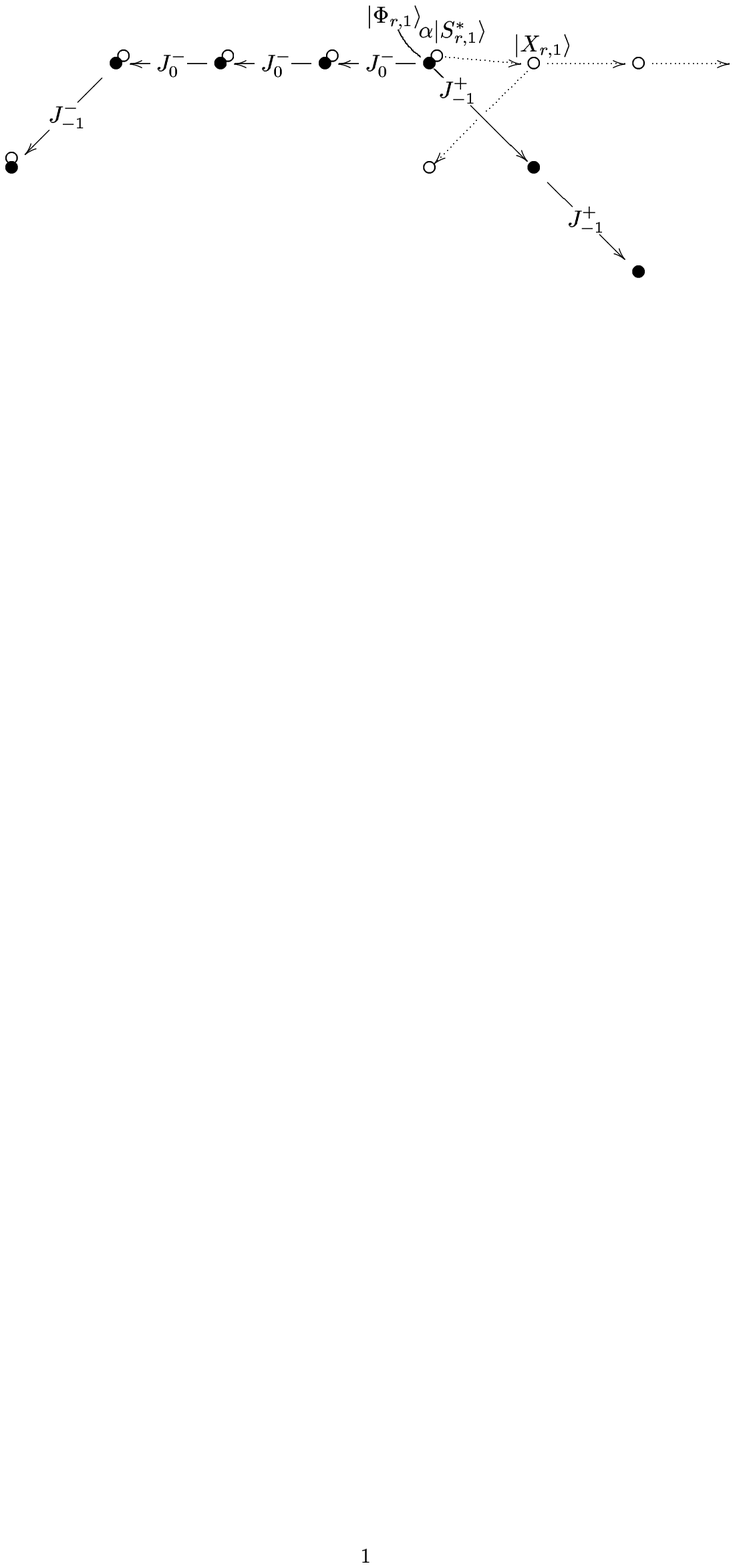}
  \end{center}
  \caption[The same as the previous Figure for
  $s\,{=}\,1$]{\captionfont{The same as Fig.\;\ref{fig:sl2} for
      $s\,{=}\,1$}.  The singular vector $(J^-_0)^r\ket{\Phi_{r,1}}$
    vanishes in the free-field realization under consideration
    ($r\,{=}\,4$ in the picture).  Dotted arrows lead to states in a
    submodule (this feature is general for all~$s$; the labels of the
    arrows, which are the same as in Fig.~\ref{fig:sl2}, are omitted
    for simplicity).  The corresponding states are therefore absent in
    the quotient.}
  \label{fig:sl2ii}
\end{figure}

In the special case where $s\,{=}\,1$, it can also be seen that the
indecomposable module does not contain any states in the grades where
$\modM_{r,s}$ contains no states because of a singular vector
vanishing.  Whenever $s\,{=}\,1$, there is the singular vector
generated by $(J^-_0)^{r}$ acting on the highest-weight vector of the
\textit{Verma} module with the same highest weight as
in~$\modM_{r,1}$; but in~$\modM_{r,1}$, we have (see
Fig.~\ref{fig:sl2ii} for~$r\,{=}\,4$)
\begin{equation}
  (J^-_0)^{r}\ket{\Phi_{r,1}}=0
\end{equation}
(for example, for $r=2$, this readily follows from the differential
equation in Eq.~\eqref{V12-diff} for the ``matter'' part $V_{21}$ of
the primary field operator).  Moreover, it is straightforward to
verify the vanishing
\begin{equation}
  (J^-_0)^{r}\ket{S^*_{r,1}}=0\,,
\end{equation}
and therefore (after taking the quotient with respect to the submodule
generated from $\ket{X_{r,1}}$), the ``logarithmic'' states exist in
only those grades where the original module~$\modM_{r,1}$ is
``populated.''

The $\tilde{L}_0$ generator is not diagonalizable on the logarithmic
module.  For the $\hSL2$ representation, the indecomposability amounts
to the action of $\tilde{J}^-_n$ ``mixing'' the $\circ$ and $\bullet$
states in the appropriate grades: each $\circ$ state is mapped into a
linear combination of the $\circ$ and $\bullet$ states in the target
grade; for $\tilde{J}^-_0$, in particular, this occurs at the top
level of the module shown in Figs.~\ref{fig:sl2} and~\ref{fig:sl2ii}.


\section{Conclusions} \label{sec:conclusions} 
We have proposed a construction that allows us to obtain logarithmic
conformal field theories 
from conventional conformal theories.  The construction amounts to
deforming the action of the operator algebra on the tensor product of
a module\,---\,either the vacuum module $V$, i.e., the algebra itself,
or some other module~$W$\,---\,with an auxiliary finite-dimensional
vector space~$K$ using a screening current~$E(z)$.  With a
screening~$\oint\!E$ thus involved in logarithmically deforming an
operator product algebra $\opaV$, the poles
$(\opepole[n][E,A])_{n\geq2}^{}$ of $E$ with $A\,{\in}\,\opaV$ (which
are irrelevant for the property of $\oint\!E$ to be a screening) start
playing the role of ``logarithmic deformation directions.'' When the
screening is fermionic, the auxiliary space $K$ is two-dimensional and
its endomorphisms are generated by two elements of the corresponding
Clifford algebra. One of the two basis elements in~$K$ (tensored with
the vacuum of the original theory) then corresponds to the new
operator~$\dd^{-1}\!E(z)$.

In a rational conformal field theory, the space $H$ of states is a
fully reducible module over the relevant chiral algebra (e.g., for
minimal models, a finite direct sum of irreducible Virasoro modules),
and there are no nontrivial extensions
\begin{align*}
  0\longrightarrow H\longrightarrow \modL\longrightarrow
  H_1\longrightarrow0
\end{align*}
of $H$ by another fully reducible module $H_1$ in the category of
diagonalizable modules.  In contrast, when one does not require
diagonalizability, nontrivial extensions do exist.  Such extensions
can indeed be obtained via our construction. In the situation with a
fermionic screening, size-$2$ Jordan cells arise, thereby providing
the state--space correspondence for a rank-$2$ logarithmic conformal
field theory. (For bosonic $E$, the deformation does not necessarily
imply the occurrence of logarithms; e.g., when $E$ is a bosonic
current ($h_E\,{=}\,1$), we have seen the deformation to reduce to the
familiar spectral flow transform associated with $E$.)

A convenient tool in the study of CFT models is often provided by some
larger algebra $\cA$ that contains the chiral algebra of the model
under consideration. (Most commonly, although not necessarily, $\cA$
is obtained via a free-field realization of the model.) In our
construction of indecomposable modules, the role of such an algebra
$\cA$ is to provide us with the spin-$1$ current $E(z)$ whose zero
mode is a screening (for example, $E(z)$ is the fermionic $\N2$
algebra generator $Q(z)$ for the $c\,{=}\,0$ Virasoro model).
Analyzing modules of the algebra $\cA$ then indicates where Jordan
cells are present in the logarithmic Virasoro modules.  When $\cA$
corresponds to a single free boson, the existence of a fermionic
screening selects the $(2,2r{+}1)$ minimal models.  But as seen in the
example of the $\hSL2$ algebra, the construction involving a fermionic
screening is not limited to the Virasoro case.

We add a word of caution. We have followed the habit in the literature
to indiscriminately use the term ``logarithmic conformal field
theory'' whenever dealing with structures that on the one hand share
crucial aspects of ordinary conformal field theories and on the other
hand display logarithmic behavior in correlation functions on
$\oC\oP^1$.  For ordinary rational CFTs, it is largely known (compare,
e.g.,~\cite{BEfr,BAki}) what is required in order to have a complete
consistent theory (in particular, with sensible fusion rules, a
modular functor, and correlation functions on arbitrary Riemann
surfaces that possess good locality and factorization properties).  In
contrast, it is as yet unclear under which circumstances a similarly
complete picture can be achieved in the logarithmic case. Some issues
that are of relevance to this problem have been addressed in the
literature (see, e.g.,~\cite{gaKa3,mila3}), but we suspect that much
more work is still needed before the term logarithmic CFT can be given
a well-defined meaning.

The construction introduced in this paper suggests that quite
generically, logarithmic theories can be related to ordinary ones by a
process of deformation and extension. In terms of the space of states,
it gives a construction of representations $\modL$ of a (rank-$2$)
logarithmic conformal field theory as an extension of a representation 
pertaining  to a standard conformal field theory with another such
representation (thus, 
both the submodule and the quotient of $\modL$ by the submodule are
diagonalizable representations from a standard conformal field
theory). We hope that continuing to study aspects of the logarithmic
extensions that we introduced will shed further light on logarithmic
conformal field theories and their appearance in physics.

\bigskip 

\noindent
\textbf{Acknowledgments}.  We are grateful to Boris Feigin and
Christoph Schweigert for interesting discussions.  This work was
supported in part by a grant from The Royal Swedish Academy of
Sciences, the INTAS grant 00-00262, and RFBR grant 99-01-00980, and
the Foundation for Promotion of Russian Science.  J.F.\ and S.H.\ are
supported in part by the Swedish Research Council.

\appendix
\section{Proof of Theorem~\ref{thm:deform}}
\label{app:proof}
The operation $\Delta_E$ defined in \eqref{Delta-E} can be
conveniently reformulated by introducing $c$-number fields $\cO_m$,
$m\,{\geq}\,0$, such that
\begin{align}
  \cO_0(z)&=\log z\,,\\
  \cO_m(z)&=\frac{1}{z^m}\,,\qquad m\,{\geq}\,1\,,
  \label{cOm}
\end{align}
with the identity operators understood in the right-hand sides.  We
then have
\begin{equation}
  \opepole[q][{\cO_m}, A]=\opepole[q][A,{\cO_m}]=0 \quad\text{for}
  \quad  q\geq1  
\end{equation}
for any operator $A$, while $\opepole[0][{\cO_m},A](z)\,{=}\,
\opepole[0][A,{\cO_m}](z)\,{=}\,A(z)/z^m$ for $m\,{\geq}\,1$, and 
similarly for~$\cO_0$.

With this notation, $\Delta_E$ reads
\begin{equation}
  \Delta_E A =
  \opepole[0][{\cO_0},{\opepole[1][E,A]}] +
  \sum_{n\geq1}\frac{(-1)^{n+1}}{n}
  \opepole[0][{\cO_{n}},{\opepole[n+1][E,A]}],
\end{equation}
and we have
\begin{align}\label{O-der-0}
  \dd^{\ell}\cO_0&=(-1)^{\ell-1}(\ell-1)!\,\cO_\ell\,,\\
  \dd^{\ell}\cO_m&=(-1)^{\ell}(m)_\ell\,\cO_{m+\ell}\,,\qquad m\,{\geq}\,1,
  \label{O-der-m}
\end{align}
where $(m)_\ell\,{=}\,m(m+1)\cdots(m+\ell-1)$ is the Pochhammer symbol.

To prove that $\Delta_E$ is a derivation, we must show that
\begin{equation}\label{must-prove}
  \opepole[q][{\Delta_E B},{C}] +
  (-1)^{EB} \opepole[q][{B},{\Delta_E C}]
  - \Delta_E\opepole[q][{B},{C}]=0\,,\qquad q\,{\geq}\,0\,.
\end{equation}
The second and the third terms in the left-hand side are easily
evaluated as
\begin{multline}\label{easy}
  (-1)^{EB}\opepole[q][{B},{\Delta_E C}]
  - \Delta_E\opepole[q][{B},{C}]={}\\
  {}=\opepole[0][{\cO_0},{\,(-1)^{EB}\opepole[q][B,{\opepole[1][E,C]}]
    -\opepole[1][E,{\opepole[q][{B},{C}]}]}]  \\
  \mbox{}\qquad + \sum_{m\geq1}\frac{(-1)^{m+1}}{m}
  \opepole[0][{\cO_{m}},{\,(-1)^{EB}\opepole[q][B,{\opepole[m+1][E,C]}]
    - \opepole[m+1][E,{\opepole[q][{B},{C}]}]}] \quad
\end{multline}
for $q\,{\ge}\,0$.  To evaluate the first term, $\opepole[q][{\Delta_E
  B},{C}]$, we note the identity~\cite{Thi}
\begin{multline}\label{left-composite}
  \opepole[q][{\opepole[p][A, B]}, C]={}\\
  {}=\sum_{\ell\geq q}(-1)^{q-\ell}\binom{p-1}{\ell-q}
  \opepole[p+q-\ell][A,{\opepole[\ell][B, C]}]
  -(-1)^{A B}\sum_{\ell\geq1}(-1)^{p-\ell}\binom{p-1}{\ell-1}
  \opepole[p+q-\ell][B, {\opepole[\ell][A,C]}].
\end{multline}
It is assumed that the binomial coefficients with a negative top entry
satisfy the same recurrence relation
\begin{equation}    
  \binom{n}{m}=\binom{n-1}{m}+\binom{n-1}{m-1}
\end{equation}
as those with $n\,{\geq}\, 0$, as well as
\begin{equation}
  \binom{n}{0}=1\,,\qquad\binom{0}{m}=\delta_{0,m}\,.
\end{equation}
This implies that
\begin{equation}
  \binom{n}{m}=(-1)^m \binom{-n+m-1}{m}\quad {\rm for}\ n\,{<}\,0
\end{equation}
and
\begin{align}\label{sum-sum}
  \sum_{\ell\ge1}
  \frac{(-1)^\ell}{\ell} \binom{m-1}{\ell-1}\binom{m-\ell}{N-\ell}&
  = - \frac1m\, \binom{m}{N} \,.
\end{align}

As a consequence of~\eqref{left-composite} and~\eqref{O-der-m}, we
obtain simple relations
\begin{align*}
  \opepole[q][{\opepole[0][{\cO_0},B]},C]
  &=\opepole[0][{\cO_0},{\opepole[q][B,C]}] +
  \sum_{\ell\geq1}\frac{(-1)^{\ell+1}}{\ell}
  \opepole[0][{\cO_{\ell}},{\opepole[q+\ell][B,C]}],
  \qquad q\,{\geq}\,1 \,,\\ 
  \opepole[q][{\opepole[0][{\cO_m},B]},C]
  &=\sum_{\ell\geq0}(-1)^{\ell}\binom{m+\ell-1}{\ell}
  \opepole[0][{\cO_{m+\ell}},{\opepole[q+\ell][B,C]}],
  \qquad m\,{\geq}\,1\,,\;\ q\,{\geq}\,1\,.
\end{align*}
Using these formulas, we now have, writing $B_n$ for
$\opepole[n][E,B]$,
\begin{multline}\label{difficult}
  \opepole[q][{\Delta_E B},{C}]
  {}=\opepole[0][{\cO_0},{\opepole[q][{B_1},C]}]
  + \sum_{m\geq1}\frac{(-1)^{m+1}}{m}
  \opepole[0][{\cO_m}, {\opepole[q][{B_{m+1}}, C]}]+{}\\
  {}+ \mathop{\sum_{m\geq1}\sum_{\ell\geq1}}_{m-\ell\geq0}
  \frac{(-1)^{m+1}}{\ell}\binom{m-1}{\ell-1}
  \opepole[0][{\cO_m}, {\opepole[q+\ell][{B_{m-\ell+1}}, C]}].\qquad
\end{multline}
Adding~\eqref{easy} and~\eqref{difficult}, we obtain
\begin{equation*}
  \opepole[q][{\Delta_E B},{C}] +
  (-1)^{EB} \opepole[q][{B},{\Delta_E C}]
  - \Delta_E\opepole[q][{B},{C}]=
  \opepole[0][{\cO_0}, {X_0}] +
  \sum_{m\geq1}\opepole[0][{\cO_m},{X_m}],
\end{equation*}
with explicit expressions for the quantities $X_m$ with $m\,{\ge}\,0$.
We must show that all $X_m$ vanish.  That $X_0$ vanishes immediately
follows using Eq.~\eqref{left-composite}. For $m\,{\ge}\,1$, we perform
simple manipulations using~\eqref{left-composite} to rewrite $X_m$ as
\begin{multline}\label{give}
  X_m=\frac{(-1)^{m+1}}{m}\sum_{\ell\geq1}(-1)^{\ell}\binom{m}{\ell}
  \opepole[m+1-\ell][E, {\opepole[q+\ell][B, C]}]\\
  {}-(-1)^{EB}\sum_{\ell\geq1}
  \frac{(-1)^{\ell}}{m}\binom{m}{\ell{-}1}
  \opepole[m+1+q-\ell][B, {\opepole[\ell][E, C]}]
  +\sum_{\ell\geq1}
  \frac{(-1)^{m+1}}{\ell}\binom{m{-}1}{\ell{-}1}
  \opepole[q+\ell][{\opepole[m-\ell+1][E, B]}, C].
\end{multline}
Here, it still remains to apply~\eqref{left-composite} to the last
term.  This is straightforward; we have
\begin{multline}
  \sum_{\ell=1}^m\frac{(-1)^{m+1}}{\ell}\binom{m-1}{\ell-1}
  \opepole[q+\ell][{\opepole[m-\ell+1][E, B]}, C]={}\\
  {}= \sum_{\ell=1}^m\sum_{n\geq q+\ell}
  \frac{(-1)^{m+1+q+n+\ell}}{\ell}
  \binom{m-1}{\ell-1}\binom{m-\ell}{n-q-\ell}
  \opepole[m+1+q-n][E, {\opepole[n][B, C]}]\\
  -(-1)^{EB}\sum_{\ell=1}^m\sum_{n\geq1}
  \frac{(-1)^{n+\ell}}{\ell}
  \binom{m-1}{\ell-1}\binom{m-\ell}{n-1}
  \opepole[m+1+q-n][B, {\opepole[n][E, C]}].\qquad
\end{multline}
We can now use identities~\eqref{sum-sum} to finally conclude that
$X_m$ vanishes for $m\,{\geq}\,1$.~\hfill\rule{1.5ex}{1.5ex}

\medskip

We note that the proof above relies heavily on the relations
\eqref{O-der-m}, but does not use the explicit
expressions~\eqref{cOm}.  As a consequence, the proof remains valid if
$\cO_m(z)$ with $m\,{\geq}\,1$ are replaced with
\begin{equation}
  \cO_m = \frac{(-1)^{m-1}}{(m-1)!}\,\dd^{m}\cO_0\,,
  \qquad m\,{\geq}\,1\,,
\end{equation}
\textit{with an arbitrary function $\cO_0$.} This observation is 
used in proving Lemma~\ref{thm:general}.

\section{The VOA setting} \label{voa}
A vertex operator algebra $\voaV$ is an algebraic structure
characterized by a collection of data and relations.  Her we only
summarize a few basic ingredients; for more details, we refer
to~\cite{FRlm,frhl,kac3}.

\noindent
The data are:
\begin{enumerate}
\item A vector space $V$ graded over $\oZ$ with finite-dimensional
  homogeneous subspaces $V_{(m)}$.

\item Two distinguished elements:
  $\Vac\,{\in}\,V_{(0)},\,L_{-2}\Vac\,{\in}\,V_{(2)}$.

\item A distinguished endomorphism $L_{-1}\,{\in}\,\End(V)$,
  satisfying $L_{-1}\Vac\,{=}\,0$\,.

\item The field--state correspondence:
  A linear mapping $Y(\cdot,z){:}\;V\,{\to}\,\End(V)[[z,z^{-1}]]$,
  with $z$ a formal variable.
\end{enumerate}
These are subject to the following relations:
\begin{enumerate}

\item $Y(\Vac,z)\,{=}\,\id_V$, and
  $Y(x,z)\Vac\,{\in}\,x{+}zV[[z]]$ for all $x\,{\in}\,V$.

\item $Y(L_{-2}\Vac,z)\,{=}\,T(z)\,{=}\,\sum_n L_{n}z^{-n-2}$, with
$[L_n,L_m]\,{=}\,(n{-}m)L_{n+m}\,{+}\,\frac{c}{12}\,(n^3{-}n)\,
\delta_{n+m,0}\,\id_V$.  
  
\item The field--state correspondence respects the grading as follows:
  $x_{-m}$ defined by $Y(x,z)=\sum_n x_{n}z^{-n-h_x}$ maps
  $V_{(l)}$ into $V_{(l+m)}$.

\item $Y(L_{-1}x,z)=\dd Y(x,z)$ for all $x\,{\in}\,V$.

\item The \textit{locality property}, stating that for any pair $x,y$
  of vectors in $V$ there is a number $N_{x,y}$ such that
  $(z\,{-}\,w)^{N_{x,y}}[Y(x,z),Y(y,w)]\,{=}\,0$.
\end{enumerate}

We also need the notion of a module over a \voa. A module~$\modW$ 
over the VOA~$\voaV\,{=}\,(V,Y,\Vac,L_{-2}\Vac)$ is defined 
(see~\cite{FRlm}) by the following data:
\begin{enumerate}
\item A graded vector space $W{=}\,\bigoplus W_{(n)}$,
  $n\,{\in}\,\oQ\,$, such that $\dim W_{(n)}\,{<}\,\infty$ and
  $W_{(n)}\,{=}\,0$ for $n$ sufficiently small.
  
\item A linear map $Y{:}\;V\,{\to}\,\End(W)\aff$ such that
  $Y(\Vac,z)\,{=}\,1_W\,{\in}\,\End(W)$ and, for all $v\,{\in}\, V$
  and $w\,{\in}\, W$, $Y(v,z)_{(n)}w\,{=}\,0$ for $n$ sufficiently
  large.
  
\item The Jacobi identity
  \begin{multline}
    z_0^{-1}\delta\!\left(\tfrac{z_1-z_2}{z_0}\right)Y(u,z_1)\,Y(v,z_2)-
    z_0^{-1}\delta\!\left(\tfrac{z_2-z_1}{-z_0}\right)Y(v,z_2)\,Y(u,z_1)
    \\
    = z_2^{-1}\delta\!\left(\tfrac{z_1-z_0}{z_2}\right)Y(Y(u,z_0)v,z_2)
    \qquad\mbox{\rm for}\quad u,v\,{\in}\, V\,.  \qquad
  \end{multline}
  
\item The space $W$ is a module over the Virasoro algebra $T(z)$, and
  $L_0w\,{=}\,nw$ for $w\,{\in}\,W_{(n)}$ and $Y(L_{-1}v,z)\,{=}\,\dd
  Y(v,z)$ for~$v\,{\in}\,V$.
  
\item There exists a linear map $Y{:}\;W\,{\to}\,\Hom(V,W)\aff$ such
  that $Y(w,z)\Vac\bigr|_{z=0}\,{=}\,w$, $w\in W$.
  
\item Rationality and associativity properties (see~\cite{FRlm} for
  details).
\end{enumerate}

\medskip

The relation of this framework to the OPA setting in~\cite{Thi} is as
follows. If $\voaV$ is a vertex operator algebra, then there is an
OPA structure on $\opaV\,{:=}\,Y(V,z)\,{\subset}\,\End(V)\aff$ and an
OPA module structure
on~$\opaW\,{:=}\,Y(W,z)\,{\subset}\,\Hom(V,W)\aff$ induced by the
operations~$\opepole[n][~,\,]$ that are read off from the operator
product expansions,
\begin{equation}
  A(z)B(w)=\sum_{n\in\oZ}\frac{\opepole[n][A,B](w)}{(z-w)^n}\,,
\end{equation}
where $A(z)\,{=}\,Y(v,z)$, $B(z)\,{=}\,Y(w,z)$ and $v\,{\in}\,V$, 
$w\,{\in}\,V\ \mbox{\rm
  or}\ W$.  These operations satisfy the well-known properties, which
can be used to axiomatically define an OPA~\cite{Thi}.  Writing
\begin{equation}
  A(z)=\sum_{n\in \oZ}z^{-n}A_{(n)}\,
\end{equation}
and using the formula
\begin{equation*}
  [A_{(n)},B(z)]=\sum_{m=1}^n\binom{n-1}{m-1}z^{n-m}
  \opepole[m][A,B](z)\,,\quad n\,{>}\,0\,,
\end{equation*}
where $[\ ,\ ]$ is the (super)commutator, these operations can be 
re-expressed as
\begin{equation}\label{bracket-new}
  \opepole[n][A,B](z)=\sum_{m=1}^n(-1)^{n-m}\binom{n-1}{m-1}
  z^{n-m}[A_{(m)},B(z)]\,,\quad n\,{>}\,0 \,.
\end{equation}
When written in this form, the definition of the products 
$\,\opepole[n][A,B]$ with $n\,{>}\,0$ can be extended to fields
$A\,{\in}\,\End(V){\aff}\;$ and $B\,{\in}\,\Hom(V,W){\aff}[\log z]$.

For $\opepole[n][~,\,]$ with nonpositive $n$, we write
\begin{align}\label{bracket-neg}
  \opepole[-n][A,B]&=\tfrac{1}{n!}\,{\boldsymbol{:}}(\dd^n
  A)B{\boldsymbol{:}}\,,
  \qquad n\geq1\,,\\
  \label{bracket-0}
  \opepole[0][A,B]&={\boldsymbol{:}}AB{\boldsymbol{:}}\,,
  \intertext{where the normal-ordered product is defined as
    in~\cite[Eq.(3.1.3)]{kac3},}
  {\boldsymbol{:}}AB{\boldsymbol{:}}(z)&
  =A(z)_+B(z)+(-1)^{AB}B(z)A(z)_-\,,
  \notag
\end{align}
with $A(z)_+$ being the part containing only creation operators and
$A(z)_-$ the one with only annihilation operators
(see~\cite[Eq.~(2.3.3)]{kac3}). We extend this definition
of~$\opepole[n][A,B]$ with~$n\,{\leq0}\,$ to fields in
$\End(W){\aff}[\log z]$.  This allows us to define, in particular,
quantities of the type ${\boldsymbol{:}}\dd^{-1}\!E A{\boldsymbol{:}}$,
where $\dd^{-1}\!E$ or $A$ involve $\log z$ in their mode
decompositions.

\end{document}